# Micro- and Nanoscale Heat Transfer in Femtosecond Laser Processing of Metals

Yuwen Zhang, D.Y. Tzou and J.K. Chen
Department of Mechanical and Aerospace Engineering
University of Missouri
Columbia, MO 65211, USA

**Table of Contents**

## ABSTRACT

Ultrafast laser material processing has received significant attention due to a growing need for the fabrication of miniaturized devices at micro- and nanoscales. The traditional phenomenological laws, such as Fourier's law of heat conduction, are challenged in the microscale regime and a hyperbolic or dual phase lag model should be employed. During ultrafast laser interaction with metal, the electrons and lattices are not in equilibrium. Various two-temperature models that can be used to describe the nonequilibrium heat transfer are presented. A semi-classical two-step heating model to investigate thermal transport in metals caused by ultrashort laser heating is also presented. The main difference between the semiclassical and the phenomenological two-temperature models is that the former includes the effects of electron drifting, which could result in significantly different electron and lattice temperature response from the latter for higher-intensity and shorter-pulse laser heating. Under higher laser fluence and/or short pulse, the lattice temperature can exceed the melting point and melting takes place. The liquid phase will be resolidified when the lattice is cooled by conducting heat away. Ultrafast melting and resolidification of the thin gold film and microparticles were investigated. At even shorter pulse width, femtosecond laser heating on metals produces a blasting force from hot electrons in the sub-picosecond domain, which exerts on the metal lattices along with the non-equilibrium heat flow. Our work that employs the parabolic two-step heating model to study the effect of the hot-electron blast in multi-layered thin metal films is also presented.

# 1. INTRODUCTION

Micro- and nanoscale heat transfer has become one of the very active areas drawn considerable attentions of many researchers due to its importance in many advanced manufacturing and materials processing. The traditional phenomenological laws, such as Fourier's law of heat conduction, are challenged in the microscale regime because (1) the characteristic lengths of the various heat carriers are comparable to each other and to the characteristic length of the system considered, and/or (2) the characteristic times of the various heat carriers are comparable to the characteristic energy excitation time (Wang and Prasad, 2000). Thus, microscale heat transfer can be referred to as heat transfer occurring on both the micro-length and micro-time scales. Microscale heat transfer finds applications in thin film (micro- length scale) as well as ultra-short pulsed laser processing (micro- time scale).

Chirped-pulse amplification was first accomplished over two decades ago (Strickland and Mourou, 1985). Development of femtosecond lasers followed, then accelerated over the past decade, because feasibility studies on metals (Momma et al., 1996; Hopkins and Sibbett, 2000), high bandgap materials (Fan et al., 2002), and biological materials (Riemann et al., 2004) revealed high precision and high-quality surfaces. Among demonstrated potentials was the capability of femtosecond lasers to produce "clean cuts," namely the removal of material with minimal collateral damage. This may be the most striking feature that has not been matched by any other material processing means. Laser intensity/fluence must be increased simultaneously, to assure sufficient processing power, as the pulse duration (heating time) for each pulse shrinks into the femtosecond domain (Yamanouchi et al., 2007). To date, with the processing power carried by each pulse entering pettawatts ($10^{15}$ W) in less than 100 femtoseconds, the femtosecond lasers remove virtually any type of material in a mere tenth of a picosecond while confining the process zone to within ten nanometers. This extremely short heating time is a key element that limits undesirable spread of the heat-affected zone, assuring minimal collateral damage on neighboring materials when clean cuts are made with femtosecond lasers.

When the laser pulse is reduced to a nanosecond ($10^{-9}$ sec) or less, the heat flux of the laser beam can be as high as $10^{12}$ W/m$^2$. For femtosecond pulse lasers, the laser intensity can even be up to $10^{21}$ W/m$^2$. Compared to long pulsed laser processing, short-pulsed laser processing enables users to precisely control the size of the heat-affected zone, the heat rate, and the interfacial velocity. Griffith *et al.* (2003) investigated femtosecond laser machining of steel and developed the femtosecond laser micro-fabrication capability in the microscale regime. Klein-Wiele *et al.* (2003) introduced a new technology of laser processing via ablation for nano-fabrication of solid materials with femtosecond laser. Heat affected zone of metals ablated with femtosecond laser was investigated by Hirayama and Obara (2003).

During laser-metal interaction, the laser energy is first deposited into electrons on the metal surface, where two competing processes occur (Hohlfeld *et al.*, 2000). One is ballistic motion of the excited electrons into deeper parts of the metal with velocity close to the Fermi velocity (~$10^6$ m/s). Another process is collision between the excited electrons and electrons around the Fermi level – an electron temperature is defined upon establishment of equilibrium among hot electrons. These hot electrons are then diffused into deeper part of the electron gas at a speed (<$10^4$ m/s) much lower than that of the ballistic motion. Meanwhile, the hot electrons are cooled by transferring their energy to the lattice through electron-phonon coupling. If the laser pulse width is shorter than the time required for the electron and lattice to achieve thermal equilibrium (thermalization time), the electrons and lattices can no longer be treated as being in thermal equilibrium (Grigoropoulos and Ye, 2000). The energy equations for the electrons and lattice must be specified separately and coupled through a coupling factor. The nonequilibrium between electrons and lattice has been observed experimentally (Eesley, 1986; Elsayed-Ali *et al.*, 1987) and can be described by the two-temperature model, which was originally proposed by Anisimov *et al.* (1974) and rigorously derived by Qiu and Tien (1993) from the Boltzmann transport equation. The nonequilibrium electron and lattice temperature model can also be derived using the dual-phase-lag model by considering lagging behavior of different energy carrier (Tzou, 1997; 2006). Chen *et al.* (2006) proposed a semi-classical two-step heating model to investigate thermal transport in metals caused by ultrashort laser heating.

Most existing two-temperature models dealt with the case that lattice temperature is well below the melting point and only pure conduction is considered. Under higher laser fluence and/or short pulse, the lattice temperature can exceed the melting point and melting takes place. The liquid phase will be resolidified when the lattice is cooled by conducting heat away. Short-pulsed laser melting of thin film involves the following three steps (Kuo and Qiu, 1996): (1) absorption of photon energy by free electrons, (2) energy transfer between the free electrons and the lattice, and (3) phase change of the lattice due to the propagation of energy. The rapid phase change phenomena induced by ultrashort pulse laser are controlled by nucleation dynamics at the interface, not by interfacial energy balance (Von Der Linde, *et al.*, 1987). The solid-liquid interface can be heated well above the melting point during a rapid melting process, in which case the solid becomes superheated. Similarly, the solid-liquid interface can be cooled far below the melting point in the rapid solidification process, in which case the liquid becomes undercooled. Both superheated solid and undercooled liquid are thermodynamically metastable states. Once phase change is triggered in a superheated solid or undercooled liquid, the solid-liquid interface can move at an extremely high velocity (on the order of 10 to $10^3$ m/s).

At even shorter pulse width, ultrafast laser heating on metals, from a mechanical point of view, disturbs the equilibrium energy state between electrons and phonons. When such a disturbance is active in the femtosecond domain, the lattices remain almost thermally undisturbed the gradient of the excessive disturbed energy over the equilibrium energy produces a net force exerting on the cold metal lattices. Magnitude of such a force increases with the power intensity of the laser, which can be sufficiently strong to destroy the metal lattices before they become noticeably heated (Falkovsky and Mishchenko, 1999). Severe deformation and the subsequent failure of the lattices, therefore, occur near the equilibrium temperature that can never be realized by conventional thermoelasticity based on lattice expansion/contraction developed in a non-uniform temperature flied. Such a hot-electron blast depends on both temperature and temperature gradient in the electron gas, resulting in pronounced effects in multi-layered metal films due to discontinuous heat transfer and load transmission across the interface.

The rest of this chapter will include a review of heat transfer models that include classical Fourier's conduction, hyperbolic, and Dual-Phase Lag (DPL) models; these are followed by several classical two-step heating models as well as a semiclassical two-step heating model. Very detailed analyses of ultrafast melting and resolidification of thin metal film and particles in micron size are discussed next. Finally, the hot electron blast, which may be the very cause for clean cuts produced by femtosecond lasers, will be discussed.

## 2. NONEQUILIBRIUM MODELS

### 2.1 Classical Heat Conduction and its Extension

### Classical Heat Conduction

The energy equation that is valid for any spatial and temporal scales can be expressed as

$$C\frac{\partial T}{\partial t} = -\nabla \cdot \mathbf{q}'' + S \tag{1}$$

where C is the heat capacity (J/m$^3$-K), $\mathbf{q}''$ is the heat flux vector, and $S$ is the internal heat source (W/m$^3$) that can be caused by volumetric heating or coupling between electron and phonon (for ultrafast heat transfer in metal). For classical engineering problems, the heat capacity is $C = \rho c_p$ and the heat flux is related to the temperature gradient by the Fourier's law:

$$\mathbf{q}'' = -k\nabla T \tag{2}$$

Substituting Eq. (2) into Eq. (1) results in the following very well known heat conduction equation:

$$\rho c_p \frac{\partial T}{\partial t} = \nabla \cdot (k \nabla T) + S \tag{3}$$

If the thermal conductivity is independent from temperature, Eq. (3) can be further simplified to

$$\frac{1}{\alpha} \frac{\partial T}{\partial t} = \nabla^2 T + \frac{S}{k} \tag{4}$$

where $\alpha = k/(\rho c_p)$ is thermal diffusivity. Equation (4) is a *parabolic* equation since the first order derivative with respect to time and the second order derivative with respect to spatial variables appeared in the equation.

**Hyperbolic Conduction Model**

The classical heat conduction theory based on Fourier's law assumes that thermal disturbance propagates with an infinite speed. As heat conduction is accomplished by successive collision of the energy carriers (phonons or electrons), the prorogation of thermal disturbance is always at a finite speed. This is particularly important for those processes involving extremely short times, cryogenic temperatures, or high heat fluxes. To account for the finite propagation speed of thermal wave, the Cattaneo-Vernotte thermal wave model can be used

$$\mathbf{q}'' + \tau \frac{\partial \mathbf{q}''}{\partial t} = -k \nabla T \tag{5}$$

where τ is the relaxation time that can be interpreted as the time scale at which intrinsic length scale of thermal diffusion ($\sqrt{\alpha t}$) is equal to the intrinsic length scale of thermal wave ($ct$) (Tzou and Li, 1993; Tzou, 1997), where $c$ is the thermal propagation speed. Thus the relaxation time is

$$\tau = \alpha / c^2 \tag{6}$$

Substituting Eq. (5) into Eq. (1) and assuming constant thermal conductivity, the following energy equation is obtained

$$\frac{1}{\alpha} \frac{\partial T}{\partial t} + \frac{\tau}{\alpha} \frac{\partial^2 T}{\partial t^2} = \nabla^2 T + \frac{1}{k} \left( S + \tau \frac{\partial S}{\partial t} \right) \tag{7}$$

which is referred to as the *hyperbolic heat conduction model* because the second order derivative of temperature with respect to time appeared. With appropriate relaxation times, Eq. (7) can be used to describe temperatures of different energy carriers – such as phonon and electron temperatures – as well be discussed later.

Mathematically, Eq. (7) is a thermal wave equation and the thermal diffusivity appears as a dumping effect of thermal propagation. In addition to the initial temperature, the initial rate of temperature, $\partial T(x,0)/\partial t$ must also be known. Consider a semi-infinite solid with initial conditions of $T = 0$ and $\partial T / \partial t = 0$ at $t = 0$. If a temperature surface temperature x = 0 is suddenly increased to $T_w$, the temperature distribution in the semi-infinite solid is shown in Fig. 1. It can be seen that a sharp wave front advances in the physical domain which separates the heat affected zone from the thermally undisturbed zone. Across the wave front, the temperature presents a finite jump while the temperature gradient presents a singularity. At $\beta = 10$ and thereafter, the wave front becomes weak and the results predicted by wave theory collapses onto those predicted by the classical

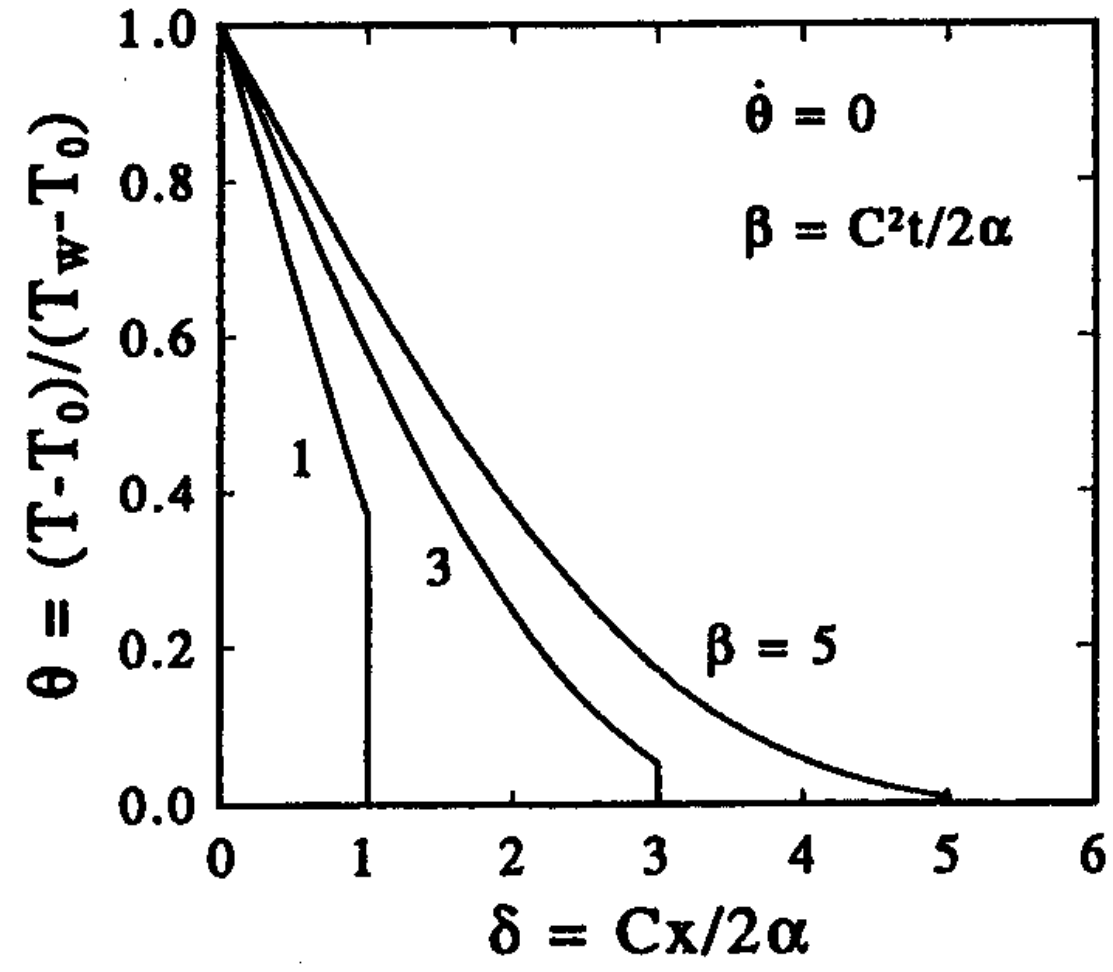

**Fig. 1** Temperature distribution in a semi-infinite solid at different times (Ozisik and Tzou, 1992)

model based on the Fourier's law.

## Dual-Phase Lag Model

Equation (5) can be viewed as the first order approximation of the following equation,

$$\mathbf{q}''(\mathbf{r}, t+\tau) = -k\nabla T(\mathbf{r}, t) \tag{8}$$

which indicates that there is a delay between the heat flux vector and the temperature gradient. For the same point in the conduction medium, the temperature gradient is established at time *t*, but the heat flux vector will be established at a later time $t+\tau$, i.e., the relaxation time, $\tau$, can be interpreted as the time delay from the onset of the temperature gradient to the heat flux vector. While the thermal wave model assumes that the temperature gradient always precedes the heat flux, Tzou (1997) proposed a dual-phase lag model that allows either the temperature gradient (cause) to precede heat flux vector (effect) or the heat flux vector (cause) to precede the temperature gradient (effect), i.e.,

$$\mathbf{q}''(\mathbf{r}, t+\tau_q) = -k\nabla T(\mathbf{r}, t+\tau_t) \tag{9}$$

where $\tau_q$ is the phase lag for the heat flux vector, while $\tau_T$ is the phase lag for the temperature gradient. If $\tau_q > \tau_T$, the *local* heat flux vector is the result of the temperature gradient at the same location but an early time. On the other hand, if $\tau_q < \tau_T$, the temperature gradient is the result of the heat flux at an early time. The first order approximation of Eq. (9) is:

$$\mathbf{q}'' + \tau_q \frac{\partial \mathbf{q}''}{\partial t} = -k\left[\nabla T + \tau_T \frac{\partial}{\partial t}(\nabla T)\right] \tag{10}$$

Substituting Eq. (10) into Eq. (1), the energy equation based on the dual-phase lag model is

$$\frac{1}{\alpha}\frac{\partial T}{\partial t}+\frac{\tau_q}{\alpha}\frac{\partial^2 T}{\partial t^2}=\nabla^2 T+\tau_T\frac{\partial}{\partial t}(\nabla^2 T)+\frac{1}{k}\left(S+\tau_q\frac{\partial S}{\partial t}\right) \tag{11}$$

which reduces to the parabolic conduction equation (3) if $\tau_q=\tau_T$. In absence of phase lag for temperature gradient ($\tau_T=0$), Eq. (11) is reduced to the hyperbolic conduction model, Eq. (7). As the values of $\tau_q$ and $\tau_T$ increase, Eq. (10) may not be accurate enough because the higher order terms were neglected. For the case that $\tau_T \ll t$, up to the first order, and $\tau_q < t$, up to the second order, Eq. (10) becomes

$$\mathbf{q}''+\tau_q\frac{\partial \mathbf{q}''}{\partial t}+\frac{\tau_q^2}{2}\frac{\partial^2 \mathbf{q}''}{\partial t^2}=-k\left[\nabla T+\tau_T\frac{\partial}{\partial t}(\nabla T)\right] \tag{12}$$

substituting Eq. (12) into Eq. (1), the heat equation with linear effect of $\tau_T$ and the second order effect of $\tau_q$ is obtained:

$$\frac{1}{\alpha}\frac{\partial T}{\partial t}+\frac{\tau_q}{\alpha}\frac{\partial^2 T}{\partial t^2}+\frac{\tau_q^2}{2\alpha}\frac{\partial^3 T}{\partial t^3}=\nabla^2 T+\tau_T\frac{\partial}{\partial t}(\nabla^2 T)+\frac{1}{k}\left(S+\tau_q\frac{\partial S}{\partial t}+\frac{\tau_q^2}{2}\frac{\partial^2 S}{\partial t^2}\right) \tag{13}$$

where the third order derivative of temperature with respect to time appeared. In addition, the second order derivative of the heat source also appeared in the heat equation.

Figure 2 shows a typical response of thermal lagging with $\beta=t/\tau_q$ and $B=\tau_T/\tau_q=100$, which are typical values for metals. It can be seen that the dual-phase lag model with the $\tau_q^2$ effect displays a wave behavior in heat propagation. The wave front that is located at $x=\left(\sqrt{2\alpha\tau_T/\tau_q}\right)t$, separates the heat affected zone from the thermally unaffected zone. The linear DPL model, on the other hand, exhibits high-order diffusion behavior. While the DPL-$\tau_q^2$ model results in high temperature than the linear DPL model in the heat affected zone, the temperature obtained from the linear DPL model is higher than that obtained by the DPL-$\tau_q^2$ model in the thermally unaffected zone. The wave propagation speed for DPL-$\tau_q^2$ model is $\sqrt{2\alpha\tau_T/\tau_q}$, which is $\sqrt{2\tau_T/\tau_q}$ times faster than the classical Cattaneo-Vernotte (CV) wave. Obviously, the speed of DPL-$\tau_q^2$ model will be slower if $\tau_T/\tau_q < 0.5$. The result of classical Fourier conduction can be obtained by setting $\tau_q = \tau_T$. The temperature obtained from the classical Fourier conduction is between the results obtained from the CV wave and DPL models.

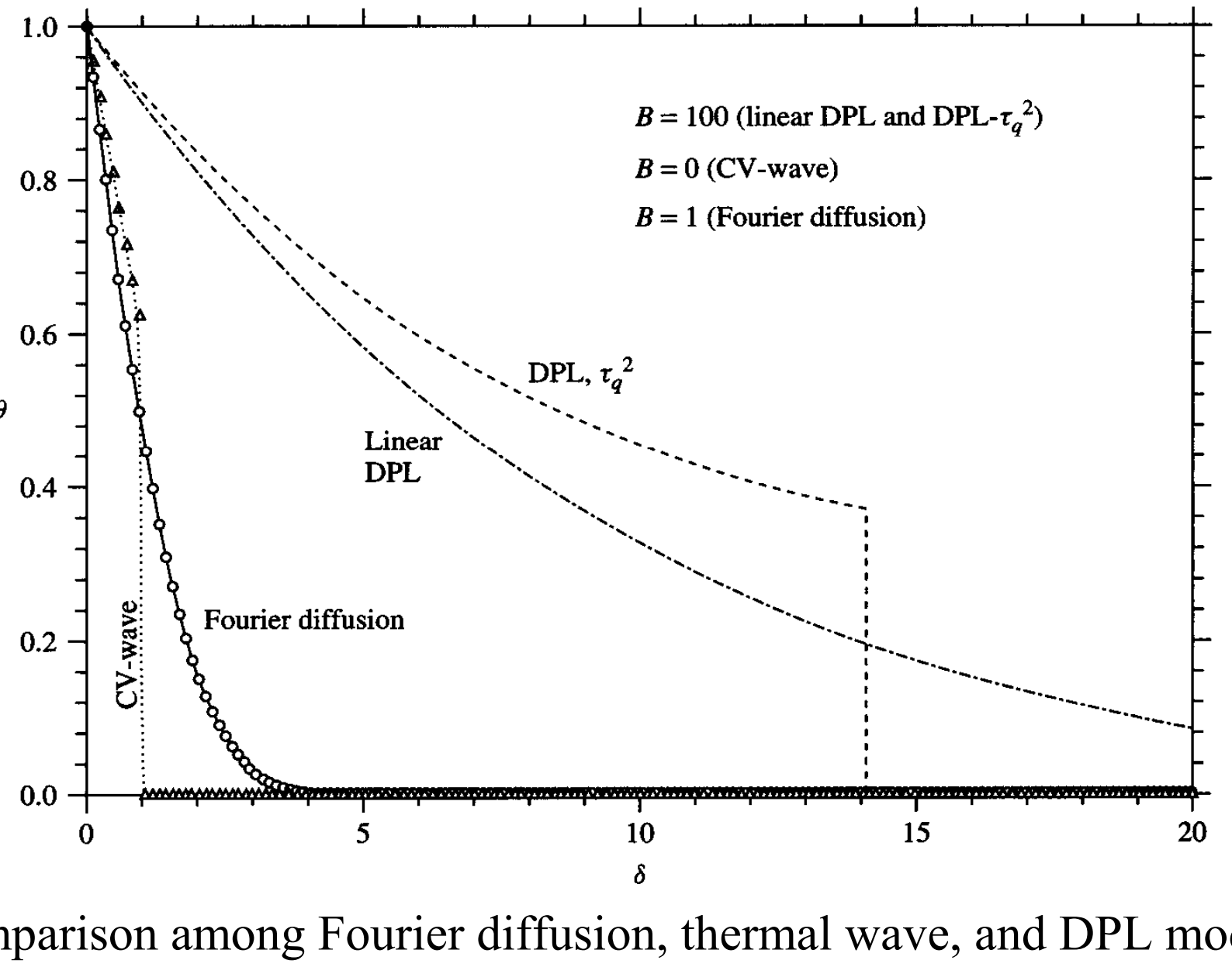

**Fig. 2** Comparison among Fourier diffusion, thermal wave, and DPL models (Tzou, 2006)

## 2.2 Two-Step Models

For ultrafast laser materials processing, the laser pulse duration is much shorter than the time it takes for the electron and lattice to reach to thermal equilibrium. Various two-step heating models can be employed to describe such nonequilibrium phenomena between the electrons and lattices.

### Parabolic Two-Step Model

Assuming heat conduction in the electron can be described by the Fourier's law and neglecting heat conduction in the lattice, the energy equations of the free electrons and lattices (phonons) are

$$C_e \frac{\partial T_e}{\partial t} = \nabla \cdot \left(k_e \nabla T_e\right) - G\left(T_e - T_l\right) + S \tag{14}$$

$$C_l \frac{\partial T_l}{\partial t} = G\left(T_e - T_l\right) \tag{15}$$

where the volumetric heat capacity of lattice is $C_l = \rho c_{\text{p}}$, and the volumetric heat capacity of electrons is

$$C_e = \frac{\pi^2 n_e k_B}{2\mu_F} T_e = B_e T_e \tag{16}$$

where $n_e$ is the number density of electrons, $k_B$ is the Boltzmann constant, and $\mu_F$ is Fermi energy. Equation (16) indicates that the volumetric heat capacity of the electron is proportional to the electron temperature. It should be noted that the volumetric heat capacity of electron is much less than that of the lattice even at very high electron temperature. At nonequilibrium condition, thermal conductivity of the electrons depends on the temperatures of both electrons and lattice, i.e.,

$$k_e = k_{eq}\left(\frac{T_e}{T_l}\right) \tag{17}$$

where $k_{eq}(T)$ is the thermal conductivity of the electron when the electrons and lattice are in thermal equilibrium. The electron-phonon coupling factor, $G$ in $\text{W/m}^3\text{K}$, is a new property in microscale heat transfer, which describes the time-rate of exchange of thermal energy between electrons and phonons per unit volume. It can be estimated by

$$G = \frac{9}{16}\frac{n_e k_B^2 T_D^2 v_F}{\Lambda(T_l) T_l \mu_F} \tag{18}$$

where $T_D$ is Debye temperature, $v_F$ is Fermi velocity, and $\Lambda$ is the electron mean free path. Neglecting conduction in the lattice is justified by the fact that the thermal conductivity of the lattice is two orders of magnitude smaller than that of the free electrons (Klemens and Williams, 1986). The heat conduction model represented by Eqs. (14) and (15) is referred to as a *parabolic two-step model* because Fourier's law was used to describe heat conduction in the electron gas.

If the thermophysical properties of electrons and lattices can be assumed to be constant, Eqs. (14) and (15) can be combined to eliminate the electron temperature and obtain the following equation for the lattice temperature

$$\frac{C_e + C_l}{k_e}\frac{\partial T}{\partial t} + \frac{C_e C_l}{G k_e}\frac{\partial^2 T}{\partial t^2} = \nabla^2 T_l + \frac{C_l}{G}\frac{\partial}{\partial t}(\nabla^2 T) + \frac{S}{k_e} \tag{19}$$

where the subscript "*l*" for lattice has been dropped for notation. Comparing Eq. (19) with the energy equation for the dual-phase lag model, Eq. (11), it is apparent that they have almost identical form except the partial derivative of heat source with respective to time is not present in Eq. (19). The thermophysical properties in the dual-phase lag model is related to the properties appeared in the two-temperature model by

$$k = k_e,\ \ \alpha = \frac{k_e}{C_e + C_l},\ \ \tau_T = \frac{C_l}{G},\ \ \tau_q = \frac{C_e C_l}{G(C_e + C_l)} \tag{20}$$

The ratio of two phase-lag times is

$$\frac{\tau_T}{\tau_q} = \frac{C_e + C_l}{C_e} = 1 + \frac{C_l}{C_e} \tag{21}$$

which indicates that $\tau_T$ is always greater than $\tau_q$.

### Hyperbolic Two-Step Model

If we considered the hyperbolic effect on the conduction in the electron gas, the energy equation for the electron gas is

$$C_e \frac{\partial T_e}{\partial t} = -\nabla \cdot \mathbf{q}'' - G\left(T_e - T_l\right) + S \tag{22}$$

where

$$\mathbf{q}'' + \tau_e \frac{\partial \mathbf{q}''}{\partial t} = -k_e \nabla T_e \tag{23}$$

while the energy equation for the lattice is still Eq. (15). Equations (22) and (23) can be combined to yield

$$\begin{aligned} C_e \frac{\partial T_e}{\partial t} + C_e \tau_e \frac{\partial^2 T_e}{\partial t^2} &= \nabla \cdot (k_e \nabla T_e) \\ -G\left(T_e - T_l\right) - \tau_e \frac{\partial}{\partial t}\left[G\left(T_e - T_l\right)\right] &+ S + \tau_e \frac{\partial S}{\partial t} \end{aligned} \tag{24}$$

The conduction model represented by Eqs. (24) and (15) is referred to as a *hyperbolic two-step model.* Qiu and Tien (1993) simulated picosecond laser-metal interaction using different models. Figure 3 shows comparison of the front surface electron temperature change of a 0.1 μm- thick gold film subject to a 96 fs-laser pulse irradiation. It can be seen that the parabolic one-step (POS) model (classical Fourier's conduction) and hyperbolic one-step (HOS) model (C-V wave model) significantly over predict the electron temperature change. The parabolic two-step (PTS) model predicted the general temperature response but it failed to predict the finite speed of energy propagation. Therefore, they suggested that the hyperbolic two-step model can provide better accuracy for ultrafast laser interaction with metal.

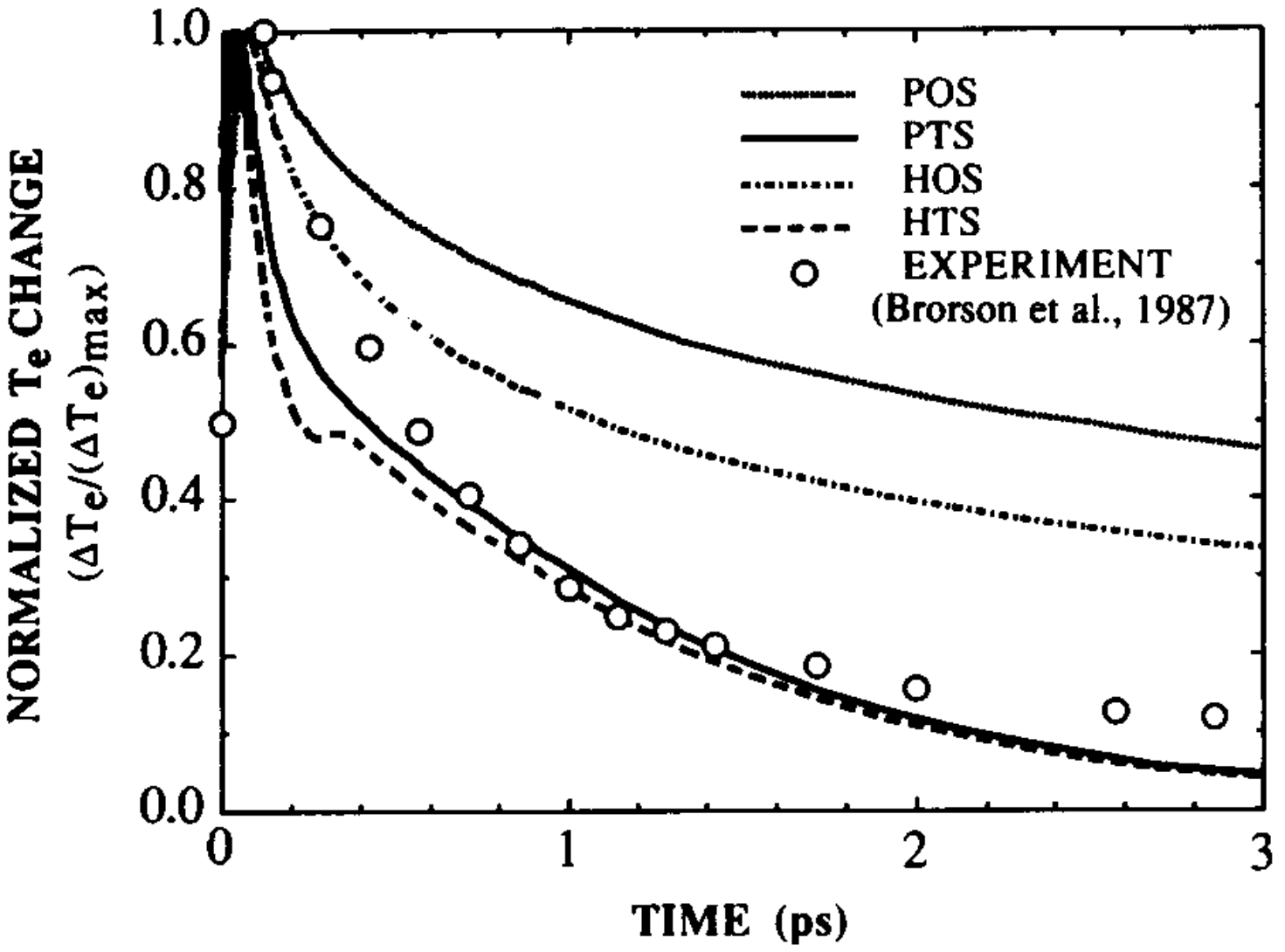

**Fig. 3** Comparison of electron temperature change (Qiu and Tien, 1993)

**Dual-Parabolic Two-Step Model**

The contribution of heat conduction in phonon was neglected in the above two models. If it is assumed that the heat conduction in the phonons can be modeled using the classical Fourier's law, the energy equations of the lattices (phonons) are

$$C_l \frac{\partial T_l}{\partial t} = \nabla \cdot (k_l \nabla T_l) + G(T_e - T_l) \tag{25}$$

The bulk thermal conductivity of metal measured at equilibrium, $k_{eq}$, is the sum of electron thermal conductivity, $k_e$, and the lattice thermal conductivity, $k_l$. Since the mechanism for heat conduction in metal is diffusion of free electron, $k_e$ is usually dominate. For gold, $k_e$ is 99% of $k_{eq}$, while $k_l$ only contributes to 1% of $k_{eq}$ (Klemens and Williams, 1986). Since both Eqs. (14) and (25) are parabolic, this model is referred to as *dual-parabolic two-step model.* For the case that phonon temperature gradient is significant, inclusion of conduction in phonon is essential.

**Dual-Hyperbolic Two-Step Model**

For the case that heat conduction in both electrons and phonons need to be considered using hyperbolic model, the energy equation for the lattice becomes

$$C_l \frac{\partial T_l}{\partial t} = -\nabla \cdot \mathbf{q}_l'' + G\left(T_e - T_l\right) \tag{26}$$

where

$$\mathbf{q}_l'' + \tau_l \frac{\partial \mathbf{q}_l''}{\partial t} = -k_l \nabla T_l \tag{27}$$

Combining (26) and (27) to eliminate $\mathbf{q}_l''$ yield

$$C_l \frac{\partial T_l}{\partial t} + C_l \tau_l \frac{\partial^2 T_l}{\partial t^2} = \nabla \cdot (k_l \nabla T_l) + G\left(T_e - T_l\right) + \tau_l \frac{\partial}{\partial t}\left[G\left(T_e - T_l\right)\right] \tag{28}$$

Equation (28) together with Eq. (24) become governing equations for the *dual-hyperbolic two-step* model. Chen and Beraun (2001) applied the dual-hyperbolic two-step model to simulate ultrashort laser pulse interactions with metal film. They found that the electron temperatures obtained from the dual-hyperbolic model and the hyperbolic model are very close. However, the lattice temperatures obtained from the two models differ significantly.

# 3 SEMICLASSICAL TWO-STEP HEATING MODEL

In a non-uniformly heated conductor in the absence of current there arises an electric field (Ashcroft and Mermin, 1976). On the other hand, substantial electron heating will occur if electric field and carrier gradient are excessively large (Snowden, 1986). When an opaque solid medium is heated by ultrashort laser pulses, very high carrier gradient and, in turn, a strong electric field can be generated due to the fact that interaction between photons and electrons and the subsequent non-equilibrium thermal transport are essentially confined to a very small region where the laser light penetrates. There is a need to investigate whether the electric field and carrier gradient caused by ultrashort laser pulses can influence thermal response of electrons and the lattice.

In this section, a semiclassical two-step heating model that includes the conservation equations of number density, momentum and energy for electron subsystem is presented. These equations are derived from the first three moments of the Boltzmann transport equation. For the phonon subsystem the thermal transport equation remains the same as that used in phenomenological two-temperature (2T) models (Anisimov *et al.*, 1974; Qiu and Tien, 1993; Chen and Beraun, 2001). The main difference between the semiclassical and the phenomenological 2T models is that the former includes the effects of the electron drift velocity resulting from non-uniform electric temperature field.

## 3.1 Model Formulation

Let $f(\mathbf{r}, \mathbf{u}, t)$ be a distribution function of the free electrons at time $t$ in the phase space with position vector ($\mathbf{r}$) and velocity vector ($\mathbf{u}$). If the Lorentz force due to magnetic field is negligible, the Vlasov equation becomes (Kruer, 1988)

$$\frac{df}{dt} = \frac{\partial f}{\partial t} + \mathbf{u} \cdot \nabla_r f + \frac{e}{m}\mathbf{E} \cdot \nabla_u f = \left(\frac{\partial f}{\partial t}\right)_c \tag{29}$$

where the del operators $\nabla_r = \partial / \partial x_i$ and $\nabla_u = \partial / \partial u_i$ ($i$ = 1, 2, 3 refers to the direction); $e\mathbf{E}/m$ is the Lorentz force resulting from electric field $\mathbf{E}$ with $e$ and $m$ denoting the charge and mass of an electron. The term $(\partial f/\partial t)_c$ is the time-rate of change of $f$ due to the electron collision.

To derive the conservation equations of number density, momentum and energy for the electron subsystem, electron number density $n(\mathbf{r},t)$, mean velocity vector $\mathbf{v}(\mathbf{r},t)$, kinetic pressure dyad $\tilde{P}(\mathbf{r},t)$, and energy flux vector $\mathbf{q}''(\mathbf{r},t)$ are defined below (Kruer, 1988):

$$n(\mathbf{r},t) = \int f(\mathbf{r},\mathbf{u},t) d\mathbf{u} \tag{30}$$

$$\mathbf{v}(\mathbf{r},t) = \frac{1}{n}\int \mathbf{u} f(\mathbf{r},\mathbf{u},t) d\mathbf{u} \tag{31}$$

$$\tilde{P}(\mathbf{r},t) = m\int (\mathbf{u}-\mathbf{v})(\mathbf{u}-\mathbf{v}) f(\mathbf{r},\mathbf{u},t) d\mathbf{u} \tag{32}$$

$$\mathbf{q}''(\mathbf{r},t) = \frac{m}{2}\int (\mathbf{u}-\mathbf{v})\cdot(\mathbf{u}-\mathbf{v})(\mathbf{u}-\mathbf{v}) f(\mathbf{r},\mathbf{u},t) d\mathbf{u} \tag{33}$$

The continuity equation can be derived by integrating Eq. (29) over velocities

$$\int \left(\frac{\partial f}{\partial t} + \mathbf{u} \cdot \nabla_r f + \frac{e}{m}\mathbf{E} \cdot \nabla_u f\right) d\mathbf{u} = \int \left(\frac{\partial f}{\partial t}\right)_c d\mathbf{u} \tag{34}$$

Since $\mathbf{r}$ and $\mathbf{u}$ are independent, Eq. (34) can be re-expressed as

$$\frac{\partial n}{\partial t} + \nabla_r \cdot (n\mathbf{v}) = \left(\frac{\partial n}{\partial t}\right)_c \tag{35}$$

In the above simplification the fact that $f \to 0$ as $|\mathbf{v}| \to \infty$ and the condition that $\mathbf{E}$ is independent of $\mathbf{u}$ are used.

Multiplying Eq. (29) by the velocity $\mathbf{u}$ and integrating the equation over the velocity space results in the equation of the conservation of momentum

$$m\frac{\partial \mathbf{v}}{\partial t} + m\mathbf{v} \cdot \nabla_r \mathbf{v} - e\mathbf{E} + \frac{1}{n}\nabla_r \cdot \tilde{P} = m\left(\frac{\partial \mathbf{v}}{\partial t}\right)_c \tag{36}$$

In absence of current the electric field $\mathbf{E} = \beta \nabla T_e$ with $\beta = -1.42\times 10^{-4}\left(T_e / T_F\right)$ V/K for free electrons (Ashcroft and Mermin,, 1976). Assume that the pressure is isotropic, i.e., $\tilde{P} = p\tilde{I} = nk_B T_e \tilde{I}$, where $k_B$ is the Boltzmann constant and $\tilde{I}$ the unit dyad. The collision term $m\left(\partial \overline{v} / \partial t\right)_c$ is approximated as (Snowden, 1986)

$$m\left(\frac{\partial \mathbf{v}}{\partial t}\right)_c = -m\frac{\mathbf{v}}{\tau_k} \tag{37}$$

where $\tau_k = m\mu_o T_l/eT_e$ is the momentum relaxation time (Baccarani and Wordeman, 1985) with $\mu_o$ being the mobility of electrons at room temperature.

Similarly, the energy equation can be obtained by multiplying Eq. (29) by the kinetic energy $m\mathbf{u}\cdot\mathbf{u}/2$, averaging the result over the entire velocity space, and utilizing the continuity and momentum conservation conditions. It is as follows:

$$\frac{\partial \xi}{\partial t}+\bar{v}\cdot\nabla_r\xi+\frac{1}{n}\nabla_r\cdot(nk_BT_e\mathbf{v})+\frac{1}{n}\nabla_r\cdot\mathbf{q}_e''-e\beta\mathbf{v}\cdot\nabla T_e=\left(\frac{\partial \xi}{\partial t}\right)_c \tag{38}$$

where $\xi=m|\mathbf{v}|^2/2+3k_BT_e/2$ is the electron energy density. The collision term $(\partial\xi/\partial t)_c$, primarily due to electron-phonon scattering, is approximated as $(\partial\xi/\partial t)_c = (\xi-\xi_l)/\tau_{e-p}$ where $\xi_l$ denotes the equilibrium energy at $T_e = T_l$ and $\tau_{e-p}$ the characteristic time for electrons and phonons to reach equilibrium (Snowden, 1986). By making use of the momentum equation and including the optical-electron scattering due to laser excitation, Eq. (38) is re-written in the form

$$C_e\left(\frac{\partial T_e}{\partial t}+\mathbf{v}\cdot\nabla_rT_e+\frac{2}{3}T_e\nabla_r\cdot\mathbf{v}\right)+\nabla_r\cdot\mathbf{q}_e''=-G(T_e-T_l)+S(\mathbf{r},t) \tag{39}$$

where $G = C_e/\tau_{e-p}$ is the electron-phonon coupling factor.

### 3.2 Electron Kinetic Pressure

If one atom can give one valence electron to the free electron gas, the number density of the excited electrons at temperature $T_e << T_F$ is given by (Kittle, 1967)

$$n=\frac{1}{3}\pi^2N\frac{T_e}{T_F} \tag{40}$$

This leads to the renowned linear electron heat capacity

$$C_e=\frac{3}{2}nk_B=C_{eo}T_e \tag{41}$$

with $C_{eo}=\pi^2Nk_B/2T_F$. Substitution of Eq. (31) for $nk_B$ into the gradient of the isotropic electron kinetic pressure dyad yields

$$\nabla_r\cdot\tilde{P}=\nabla\left(\frac{2}{3}C_{eo}T_e^2\right) \tag{42}$$

The above $\nabla_r\cdot\tilde{P}$ is the so-called hot-electron blast force $\nabla\left(gC_{eo}T_e^2\right)$ that was derived by Falkovsky and Mishchenko (1999) with undetermined constant g ~ 1. In fact, this form of the hot-electron blast force is only adequate for $T_e < 0.1T_F$. A general form of the hot-electron blast force should be $2\nabla(C_eT_e)/3$. Figure 4 gives the temperature-dependent heat capacity of electrons for gold. The following relations are approximated from the result of Jiang and Tsai (2005):

$$C_e=\begin{cases} B_eT_e & T_e<T_F/\pi^2 \\ 2B_eT_e/3+C_e'/3 & T_F/\pi^2\le T_e<3T_F/\pi^2 \\ Nk_B+C_e'/3 & 3T_F/\pi^2\le T_e<T_F \\ 3Nk_B/2 & T_e\ge T_F \end{cases} \tag{43}$$

where

$$C_e'=B_eT_F/\pi^2+\frac{3Nk_B/2-B_eT_F/\pi^2}{T_F-T_F/\pi^2}\left(T_e-T_F/\pi^2\right) \tag{44}$$

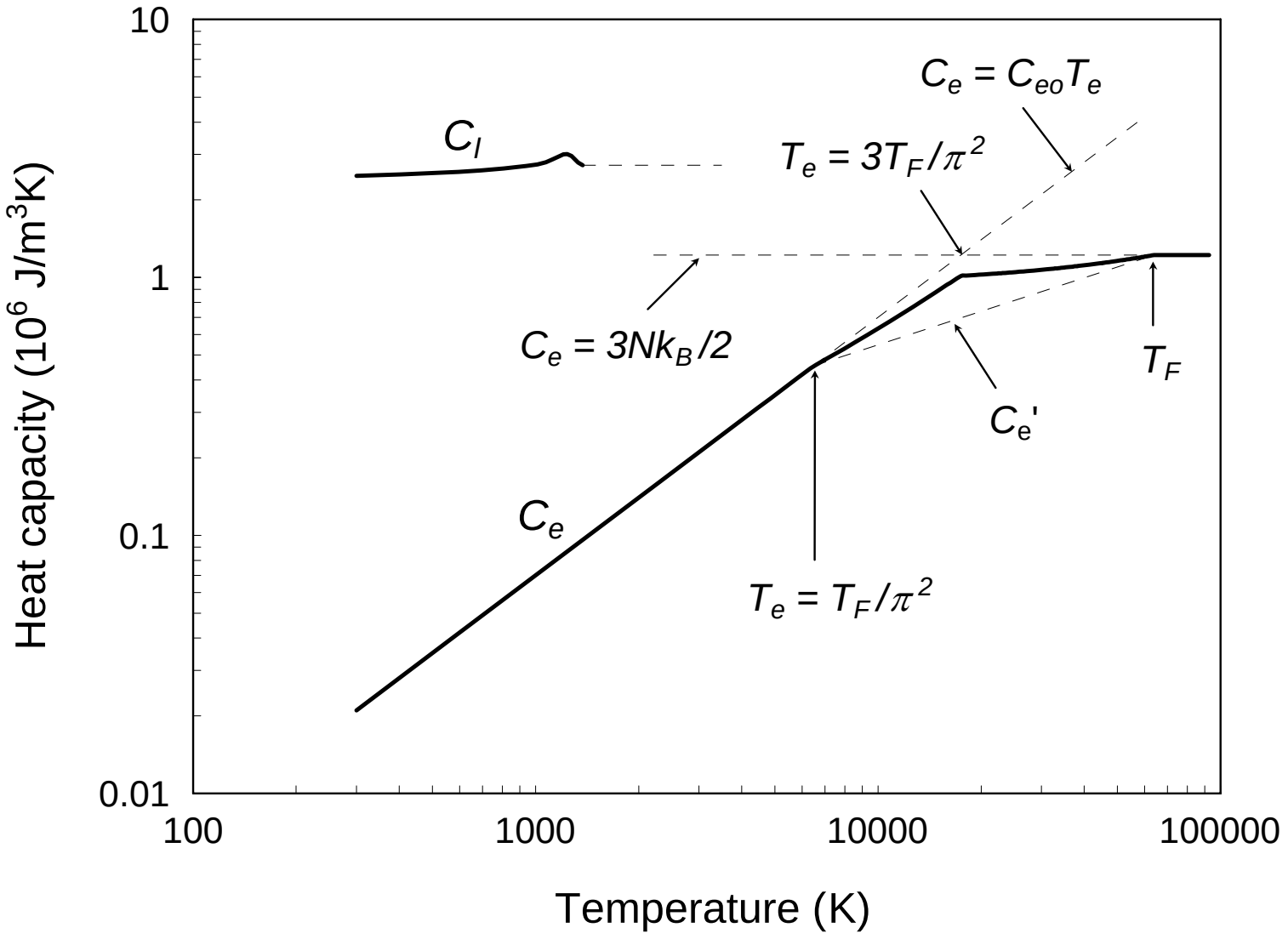

**Fig. 4** Heat capacity of gold (Chen *et al.*, 2006)

With the relationships $\nabla_r \cdot \tilde{P} = \nabla_r \left( nk_B T_e \right)$ , $\mathbf{E} = \beta \nabla T_e$ , and $\left( \partial \mathbf{v} / \partial t \right)_c$ $= eT_e \mathbf{v} / \mu_o T_l$, Eq. (36) can be re-written in terms of velocity and temperature

$$m\frac{\partial \mathbf{v}}{\partial t} + m\mathbf{v} \cdot \nabla_r \mathbf{v} + \left[ k_B \left( 1 + \frac{T_e}{C_e} \frac{\partial C_e}{\partial T_e} \right) - e\beta \right] \nabla T_e = -\frac{eT_e \mathbf{v}}{\mu_o T_l} \quad (45)$$

For low temperatures $\dfrac{T_e}{C_e}\dfrac{\partial C_e}{\partial T_e} = 1$ since $C_e = C_{eo}T_e$.

The semiclassical two-step heating model includes: (*i*) the momentum equation (45) and energy equation (39) for electron subsystem; (*ii*) the energy equation (26) for lattice subsystem; and (*iii*) the constitutive equations for heat fluxes in electrons and the lattice, Eqs. (23) and (27). For a 1D problem, for example, the above five equations are satisfied by the five unknowns, $q''_{ex}$, $T_e$, $v_x$, $Q_{lx}$ and $T_l$, while $v_x$ is excluded from the phenomenological 2T models.

### 3.3 Laser Heat Source in Two-Temperature Models

A laser heat source term widely used in 2T models is given as follows:

$$S(x,t) = -\sqrt{\frac{4\ln 2}{\pi}} \frac{(1-R)\phi}{\delta t_p} \exp\left\{ -\frac{x}{\delta} - 4\ln 2 \left[ \left( \frac{t-2t_p}{t_p} \right)^2 \right] \right\} \quad (46)$$

where $\phi$ is fluence, $R$ is surface reflectivity, $\delta$ is optical penetration depth, and $t_p$ is pulse duration. Lasing is assumed to start at $t = 0$ and end at $t = 4t_p$. Equation (46) is derived directly from the Beer's law. When it is used in the energy equation for an electron subsystem, an intrinsic assumption that the excited electrons are immediately and fully thermalized has been applied.

Immediately after a metal solid is illuminated by an ultrashort laser pulse, two competing processes take place. The photo-excited electrons move ballistically with a velocity close to the Fermi velocity. Therefore, the laser heat source used in 2T models should be modified by changing the spatial distribution of the laser heat source according to the electron ballistic motion. Meanwhile, those excited electrons continue to thermalize, through collision, into a Fermi-Dirac distribution in which the electron temperature is measurable. This also alters the time function of the laser heat source.

To incorporate the ballistic electron effect, an effective absorption depth was introduced by including an electron ballistic range to the optical penetration depth (Wellershoff *et al.*, 1999; Chen *et al.*, 2002; Ibrahim *et al.*, 2004). For instance, a range of 105 nm was assumed for gold, based on the postulation that the excited, nonequilibrium electrons move at the Fermi velocity ($V_F$) for about 100 fs (Wellershoff *et al.*, 1999). Another approach considered that the ballistic electrons scatter randomly at the Fermi velocity for a finite period of time ($t_b$) (Chen *et al.*, 2006). For a uniform 1D mesh of $q$ points, for example, the spatial function of the absorbed laser energy is modified with the initial values $\breve{S}_j = \exp[-x/\delta]$ as follows: $\breve{S}_j = (\breve{S}_{j-1} + \breve{S}_{j+1})/2$ for interior points and $\breve{S}_1 = (\breve{S}_1 + \breve{S}_2)/2$ and $\breve{S}_q = (\breve{S}_{q-1} + \breve{S}_q)/2$ for the boundary points. All $\breve{S}_j$ on the left-hand sides are updated with the previous values. The above redistribution continues for $\mathcal{N}$ times, where $\mathcal{N}$ is determined from $\mathcal{N}\Delta x/V_F = t_b$ with $\Delta x$ being grid spacing.

A system response function was proposed to account for the delayed rise time of electron temperature (Sun *et al.*, 1993; Ibrahim *et al.*, 2004). For the electrons excited at time $t'$ the delay function is given by (Ibrahim *et al.*, 2004)

$$\Psi(t,t') = u(t-t')\left\{1-\exp\left[-\left(\frac{t-t'}{t_R}\right)^2\right]\right\} \tag{47}$$

Completion of the electron thermalization, according to Eq. (47), is 0, 63.2%, and 99.99% when $t-t'$ = 0, $t_R$ and $3t_R$, respectively.

By combining both the effects of electron ballistic motion and internal thermalization, the volumetric laser heat source ($\hat{S}$) employed in a 2T model becomes

$$\hat{S}(x,t+\Delta t) = \sqrt{\frac{4\ln 2}{\pi}}\frac{(1-R)\phi}{\delta t_p}\breve{S}(x)\sum_{t'=0}^{t'=t-\Delta t}\exp\left[-4\ln 2\left(\frac{t'-2t_p}{t_p}\right)^2\right]\left[\Psi(t+\Delta t,t')-\Psi(t,t')\right] \tag{48}$$

### 3.4. Numerical Results

For numerical analysis a 1D version of the governing equations together with the following initial and boundary conditions are solved with a central difference method:

$$T_e(x,0) = T_l(x,0) = T_o;\ v_x(x,0) = 0 \tag{49}$$

$$q''_{ex}(0,t) = q''_{ex}(L,t) = 0 \tag{50}$$

where $T_o$ is set at 300 K. The thermal conductivity, $k_e$, and electron-phonon coupling factor, $G$, (Chen *et al.*, 2005) are temperature-dependent:

$$k_e = \chi \frac{(\vartheta_e^2 + 0.16)^{5/4} (\vartheta_e^2 + 0.44) \vartheta_e}{(\vartheta_e^2 + 0.092)^{1/2} (\vartheta_e^2 + \eta \vartheta_l)} \tag{51}$$

$$G = G_{RT} \left[ \frac{A_e}{B_l} \left( T_e + T_l \right) + 1 \right] \tag{52}$$

In Eq. (51) $\theta_e = T_e / T_F$ and $\theta_l = T_l / T_F$. The material properties used are as follows unless otherwise mentioned: $C_{eo}$ = 70 $\mathrm{Jm^{-3}K^{-2}}$, $k_l$ = 0, $T_F$ = 6.4 $\times 10^4$ K, $N$ = $5.9\times10^{28}$ $\mathrm{m^{-1}}$, $\mu_o$ = $4.8\times10^{-3}$ $\mathrm{m^2s^{-1}V^{-1}}$, $R$ = 0.93, δ = 15.3 nm, $A_e$ = $1.2\times10^7$ $\mathrm{K^{-2}s^{-1}}$, $B_l$ = $1.23\times10^{11}$ $\mathrm{K^{-1}s^{-1}}$, $G_{RT}$ = $2.2\times10^{16}$ $\mathrm{Wm^{-3}K^{-1}}$, χ = 353 $\mathrm{Wm^{-1}K^{-1}}$, η = 0.16, and $t_b$ = 100 fs.

Figure 5 shows the normalized change of electron temperature at the front surface ($x$ = 0) of an 80-nm gold film irradiated by a 2.8 mJ/cm$^2$, 800-nm, 150-fs laser pulse. The values of $R$, δ and $t_R$ used in this calculation are 0.967, 12.7 nm and 500 fs, respectively. The result computed with $t_R$ = 0 (i.e. all the excited electrons complete thermalization instantaneously) is also compared. The simulated maximum electron temperature is 335.9 K for the case of $t_R$ = 500 fs and 357.7 K for $t_R$ = 0, occurring at 0.988 ps and 0.358 ps, respectively. Clearly, the result computed with $t_R$ = 500 fs agrees well with the experimental data (Ibrahim *et al.*, 2004), but the case of $t_R$ = 0 fails to capture the electron temperature response. At this low level of laser fluence the difference between the semiclassical and the phenomenological 2T models is insignificant.

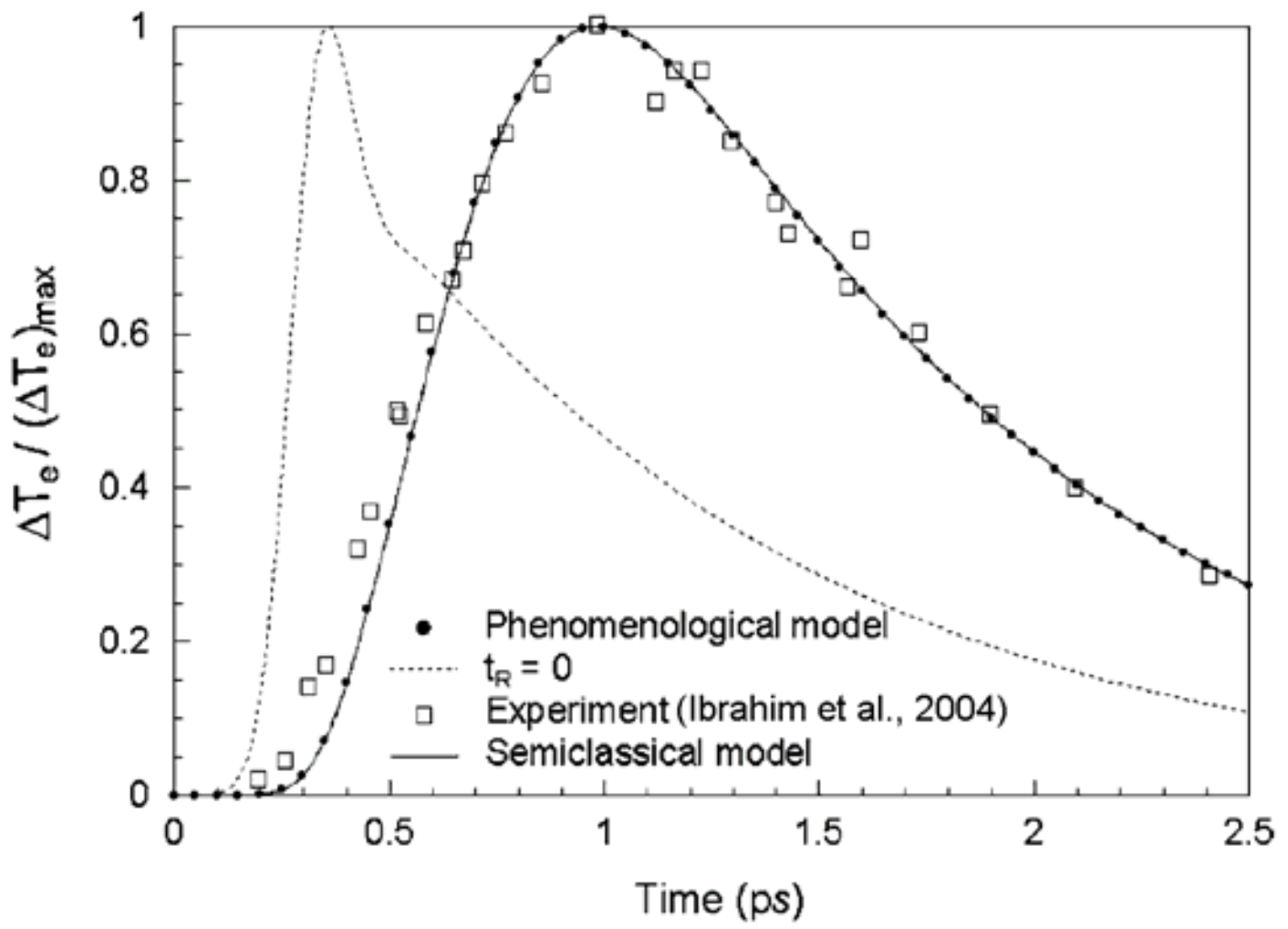

**Fig. 5** Comparison of the change in electron temperature at the front surface of an 80-nm gold film irradiated by a 2.8 mJ/cm$^2$, 800-nm, 150-fs laser pulse (Chen *et al.*, 2006)

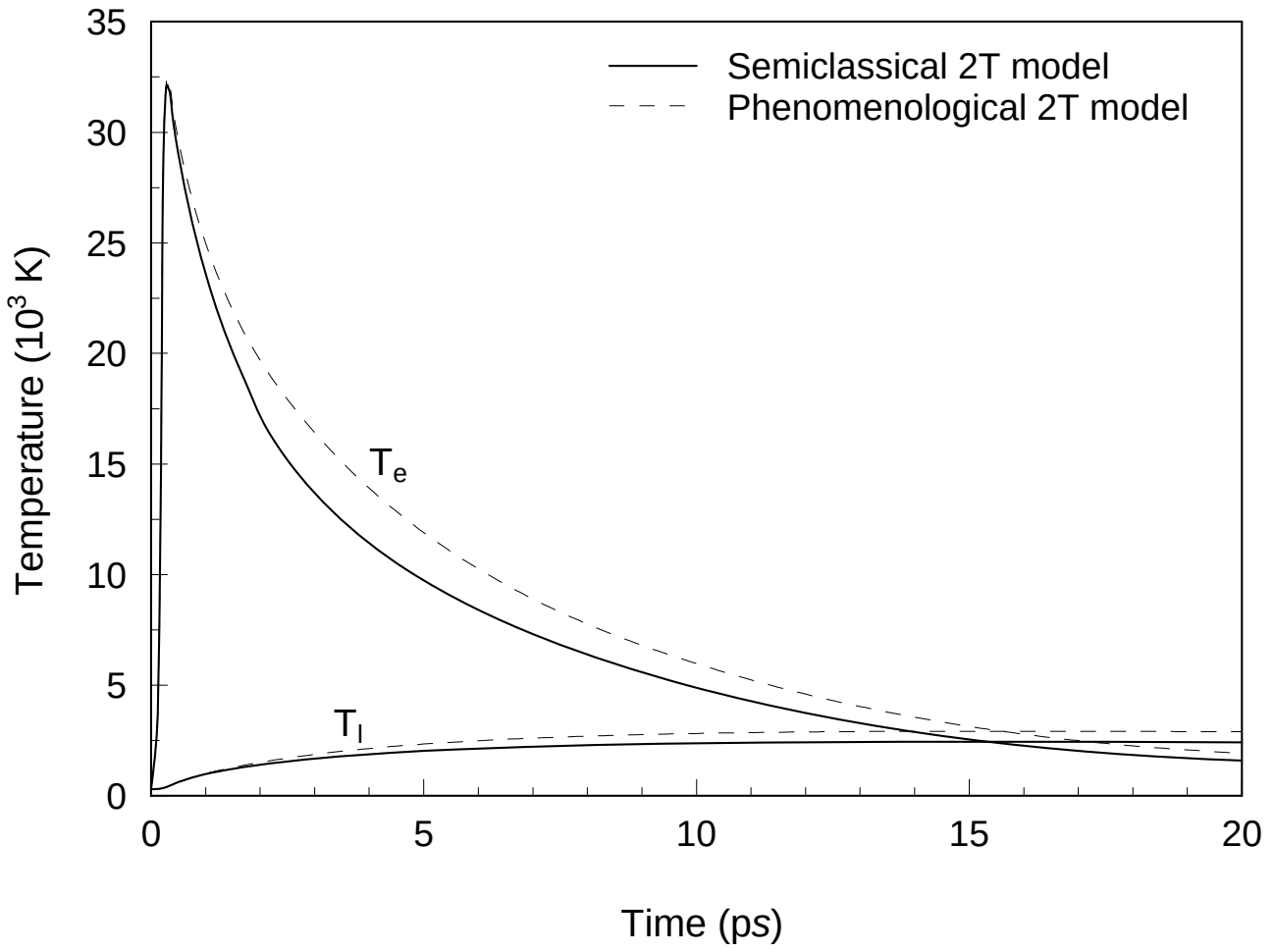

**Fig. 6** Time history of temperatures at the front surface of a 1.0-μm gold film heated with a 1.0 J/m$^2$, 560-nm, 100-fs laser pulse (Chen *et al.*, 2006).

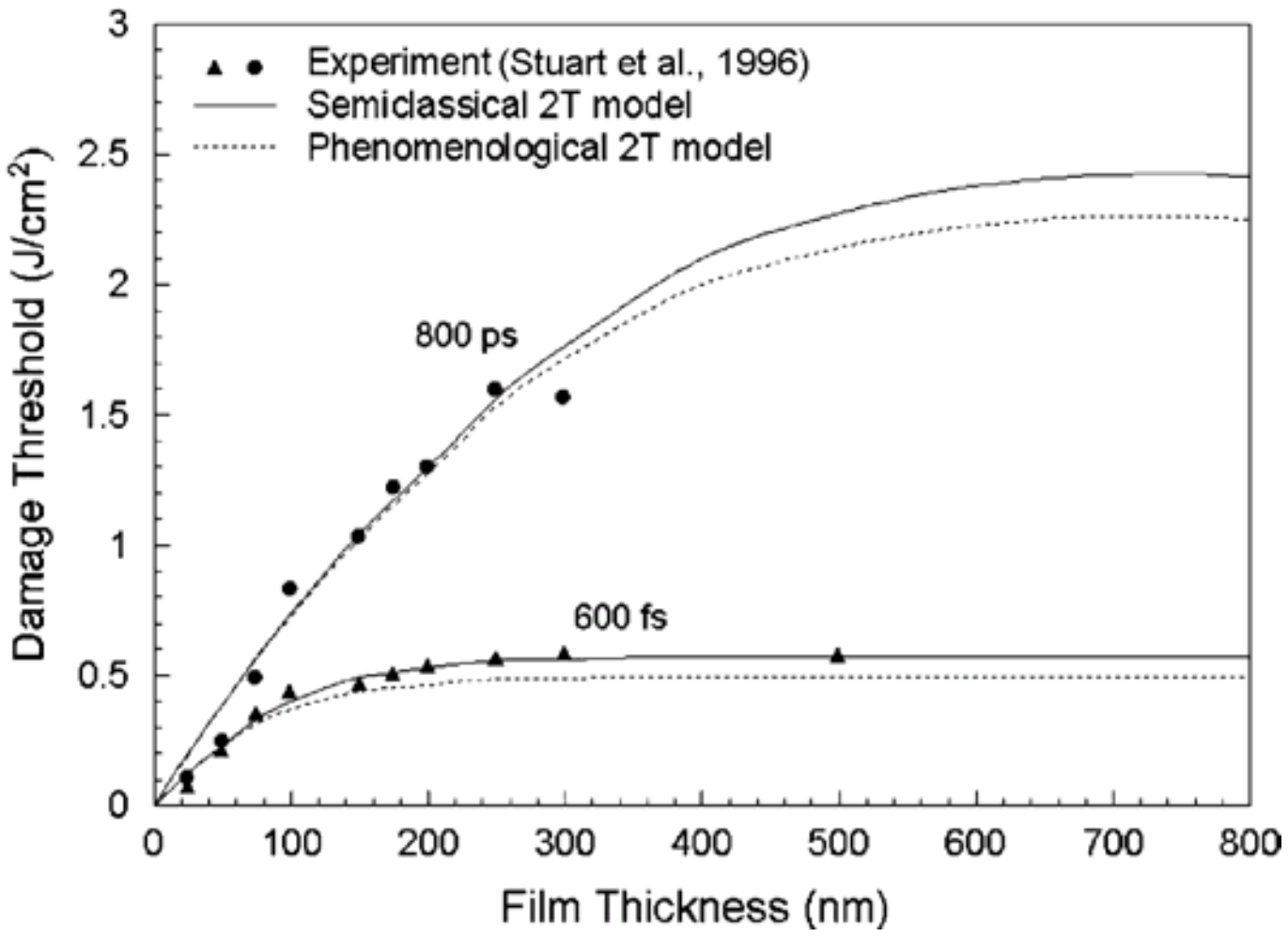

**Fig. 7** Comparison of the simulated and measured damage thresholds for a gold film heated with 600-fs and 800-ps laser pulses (Chen *et al.*, 2006)

Figure 6 shows the electron and lattice temperatures at the heated surface of a 1.0-lm gold film illuminated by a 1.0 J/cm$^2$, 630-nm, 100-fs laser pulse. At this level of laser fluence, the difference between the semiclassical and phenomenological 2T models is discernable. Further analysis shows that the difference can be immaterial for fluence 0.1 J/cm$^2$ and lower.

Figure 7 compares the simulated damage fluence threshold with the experimental data (Stuart *et al.*, 1996) for gold irradiated by two 1053-nm laser pulses of durations 600 fs and 800 ps, respectively. It is assumed here that initiation of damage occurs when the front surface completes the solid-to-liquid phase change (melting point $T_m$ = 1337 K and fusion heat $H_m$ = 6.275×10$^4$ Jkg$^{-1}$). At this wavelength the skin depth is 3 nm. The surface reflectivity used in the simulation is 0.9260 for the 600-fs pulse and 0.9546 for the 800-ps pulse. For the case of the 800-nm film heated by the 600-fs laser pulse, for example, the damage threshold obtained from the semiclassical model is about 12% higher than that obtained from the phenomenological model. The difference reduces to 7% for the longer, 800-ps pulse.

As given in the energy equation (40) for electron subsystem, the two terms involving the electron drift velocity are $\mathbf{v}\cdot\nabla_r T_e$ and $2T_e\nabla_r\cdot\mathbf{v}/3$. The numerical results show that the latter has much more impact on the change of electron temperature than the former. The difference is on the order of six (6). Accordingly, the term $\mathbf{v}\cdot\nabla_r T_e$ can be neglected without significant change of the result.

# 4. ULTRAFAST MELTING AND RESOLIDIFICATION

The models presented in the preceding two sections are for the case that the lattice temperature is below the melting point so that only pure conduction needs to be modeled. When the lattice temperature at the heating surface reaches melting point, melting takes place on the heating surface while the deeper part of the metallic target is still at room temperature (Kuo and Qiu, 1996). After the laser pulse is off, the melted layer near the heating surface will be resolidified. Such rapid melting and resolidification can find their application in selective laser sintering of metal powders using pulsed laser (Fischer *et al.*, 2002; Konrad *et al.*, 2007). During ultrashort laser pulse interaction with metal target, the solid-liquid interface velocity is controlled by nucleation kinetics, instead of energy balance (Kuo and Qiu, 1996). In this section, a fixed grid interfacial tracking method is presented to solve kinetics controlled rapid melting and resolidification during ultrashort pulse laser interaction with a free-standing metal film. Nonlinear electron heat capacity and temperature-dependent electron-lattice coupling factor on the rapid phase change are also investigated. Finally, ultrafast melting and resolidification of a submicron gold particle subject to pico- to femtosecond laser pulse are studied.

## 4.1 Interfacial Tracking Method

Many numerical models for melting and solidification of various Phase Change Materials (PCMs) have been developed. The numerical models can be divided into two groups (Vollor, 1997): *deforming grid schemes* (or strong numerical solutions) and *fixed grid schemes* (or weak numerical solutions). Deforming grid schemes transform solid and liquid phases into fixed regions by using a coordinate transformation technique. The governing equations and boundary conditions are complicated due to the transformation. However, the disadvantage of deforming grid schemes is that it requires significant amount of computational time. On the contrary, the fixed grid schemes use one set of governing equations for the whole computational domain including both liquid and solid phases, and solid-liquid interface is later determined from the temperature distribution. This simplicity makes the computation much faster than deforming grid schemes, while it still provides reasonably accurate results (Sasaguchi *et al.*, 1996). There are two main methods in the fixed grid schemes: the enthalpy method and the equivalent heat capacity method. The enthalpy method (Binet and Lacroix, 2000) can solve heat transfer in mushy zone but has difficulty with temperature oscillation, while the equivalent heat capacity method (Morgan, 1981; Hsiao, 1984) requires large enough temperature range in mushy zone to obtain converged solution. Cao and Faghri (1990) combined the advantages of both enthalpy and equivalent heat capacity methods and proposed a Temperature Transforming Model (TTM) that could also account for natural convection.

Kuo and Qiu (1996) solved rapid melting and resolidification of gold film subject to picosecond laser heating. The energy equation for electrons was solved using a semi-implicit Crank-Nicholson scheme, while the energy equation and phase change for lattices was solved using an explicit enthalpy model. While explicit scheme is very easy to apply, a very small time step is required to ensure stability. Development of an implicit scheme for enthalpy model that can outperform explicit scheme is numerically challenging (Voller, 1997). The temperature transforming model (Cao and Faghri, 1990) and the enthalpy linearization model (Swaminathan and Voller, 1993) are two very efficient models that can be easily discretized using implicit scheme. These two models assume phase change occurs in a range of temperature near melting point, which prevents inclusion of superheat in solid in melting stage and undercooling of liquid in the solidification stage. The interfacial tracking method proposed by Zhang and Chen (2008a) can be used to accurately predict the kinetics controlled melting and resolidification processes.

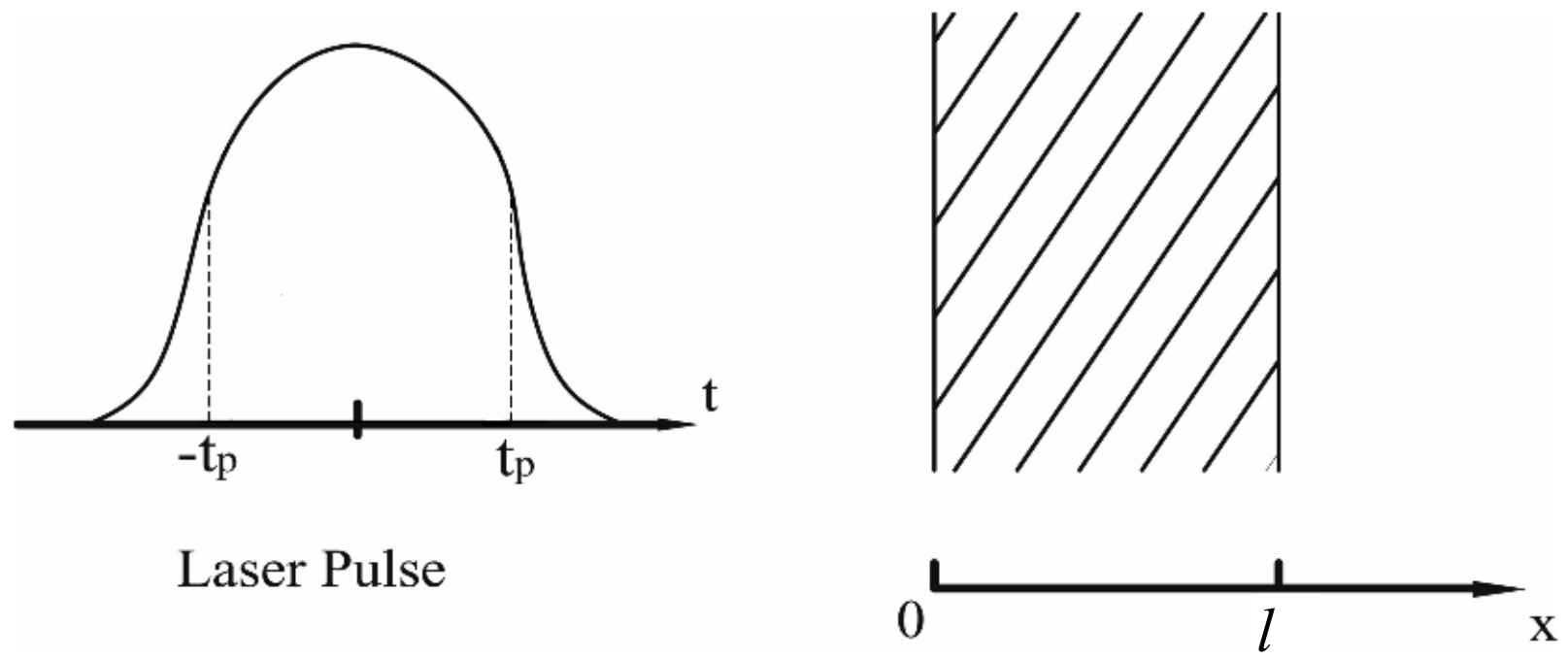

**Fig. 8** Physical model for laser melting of thin film (Zhang and Chen, 2008a)

The physical model of the problem under consideration is shown in Fig. 8. A metal film with a thickness of $L$ and an initial temperature of $T_i$ is subjected to a temporal Gaussian laser pulse with a FWHM (Full Width at Half Maximum) pulse width of $t_p$ and fluence of $J$ (J/m$^2$) from the left surface ($x$ = 0). The problem can be approximated to be one-dimensional because the radius of the laser beam is significantly larger than the metal film thickness. The dual parabolic two-step model will be used to describe the ultrafast laser-metal thin film interaction. The energy equations of the free electrons and lattice can be obtained by simplifying Eqs. (14) and (25), i.e.,

$$C_e \frac{\partial T_e}{\partial t} = \frac{\partial}{\partial x}\left(k_e \frac{\partial T_e}{\partial x}\right) - G\left(T_e - T_l\right) + S \tag{53}$$

$$C_l \frac{\partial T_l}{\partial t} = \frac{\partial}{\partial x}\left(k_l \frac{\partial T_l}{\partial x}\right) + G\left(T_e - T_l\right) \tag{54}$$

Equation (53) is valid in the entire computational domain. On the contrary, Eq. (54) is valid in both solid and liquid phases but not at the solid-liquid interface. The heat capacity of electron is proportional to the electron temperature as indicated by Eq. (16). The thermal conductivity of the electrons is obtained from Eq. (17). The lattice thermal conductivity, $k_\ell$, is taken as 1% of the thermal conductivity of bulk metal, $k_{eq}$, since the mechanism of heat conduction in metal is mainly by electrons , i.e.,

$$k_l = 0.01 k_{eq} \tag{55}$$

The source terms in Eq. (53) can be described by the following equation:

$$S = 0.94 \frac{1-R}{t_p \delta} J \exp\left[\frac{x}{\delta} - 2.77\left(\frac{t}{t_p}\right)^2\right] \tag{56}$$

where $R$ is reflectivity of the thin film, $t_p$ is laser pulse duration (s), $\delta$ is the optical penetration depth (m), and $J$ is laser pulse fluence (J/m$^2$).

The energy balance at the solid-liquid interface is (Faghri and Zhang, 2006)

$$k_{l,s} \frac{\partial T_{l,s}}{\partial x} - k_{l,\ell} \frac{\partial T_{l,\ell}}{\partial x} = \rho_\ell h_m u_s \qquad x = s(t) \tag{57}$$

where the additional interfacial velocity due to the density change during melting and solidification has been considered.

For the conventional melting process, the velocity of the solid-liquid interface is obtained by energy balance as specified by Eq. (57). However, this is not the case for rapid melting/solidification processes, because the velocity of the interface is dominated by nucleation dynamics. For short-pulsed laser melting of gold, the velocity of the solid-liquid interface is described by (Kuo and Qiu, 1996)

$$u_s = V_0 \left[1 - \exp\left(-\frac{h_m}{R_g T_m} \frac{T_{l,I} - T_m}{T_{l,I}}\right)\right] \tag{58}$$

where $V_0$ is the maximum interface velocity, $R_g$ is the gas constant for the metal, and $T_{l,I}$ is the interfacial temperature. The interfacial temperature, $T_{l,I}$, is higher than melting point, $T_m$, during melting and lower than melting point during solidification.

The time $t$ = 0 is defined as the time when the peak of a laser pulse reaches the film surface. Therefore, the initial conditions of the problem are

$$T_e(x, -2t_p) = T_l(x, -2t_p) = T_i \tag{59}$$

The boundary conditions of the problem can be specified by assuming that the heat loss from the film surface can be neglected, i.e.,

$$\left.\frac{\partial T_e}{\partial x}\right|_{x=0} = \left.\frac{\partial T_e}{\partial x}\right|_{x=l} = \left.\frac{\partial T_l}{\partial x}\right|_{x=0} = \left.\frac{\partial T_l}{\partial x}\right|_{x=l} = 0 \tag{60}$$

The computational domain (0, $L$) is discretized with non-uniform grids. The implicit finite difference equations are obtained by integrating Eqs. (53) and (54) in each control volume and time step (Pantankar, 1980). For the control volume that contains solid-liquid interface, the lattice temperature, $T_{l,P}$, is numerically set as interfacial temperature, $T_{l,I}$. The energy equation for lattice, Eq. (54) is valid for solid and liquid phase only, but not at the interface. The energy equation for the control volume that contains solid-liquid interface (see Fig. 9) can be written in the enthalpy form:

$$\frac{\partial H_l}{\partial t} = \frac{\partial}{\partial x}\left(k_l \frac{\partial T_l}{\partial x}\right) + G\left(T_e - T_l\right) \quad (61)$$

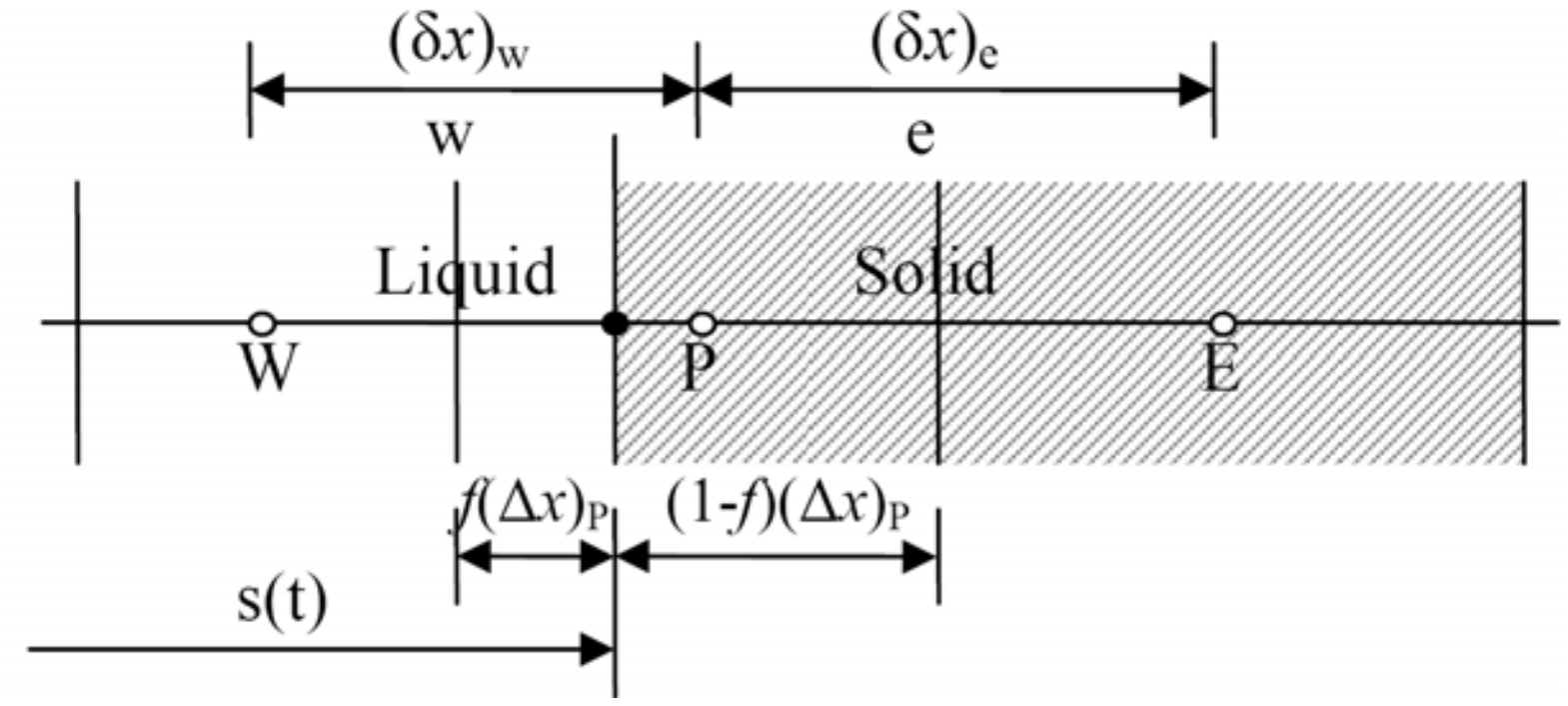

**Fig. 9** Grid system for control volume with interface

The volumetric enthalpy can be expressed as

$$H_l = \int_0^{T_{l,I}} C_{l,s}(T_l)dT_l + f\rho_\ell h_m \quad (62)$$

where the first term is enthalpy of solid phase at the interfacial temperature, and the second term is the latent heat due to partial melting. Substituting Eq. (62) into Eq. (61), one obtains

$$C_{l,s}(T_{l,I})\frac{\partial T_{l,I}}{\partial t} + \rho_\ell h_m \frac{\partial f}{\partial t} = \frac{\partial}{\partial x}\left(k_l \frac{\partial T_l}{\partial x}\right) + G\left(T_e - T_l\right) \quad (63)$$

which is applicable in the control volume that contains solid-liquid interface. The liquid fraction, $f$, is related to the location of the solid-liquid interface by (see Fig. 9).

$$\frac{\partial f}{\partial t} = \frac{1}{(\Delta x)_P}\frac{ds}{dt} = \frac{u_s}{(\Delta x)_P} \quad (64)$$

where $(\Delta x)_P$ is the size of the control volume $P$, and $s$ is the solid-liquid interfacial location. Substituting Eq. (64) into Eq. (63) and integrating the resulting equation in the control volume that contains solid-liquid interface, the interfacial velocity is obtained as

$$u_s = \frac{1}{\rho_\ell h_m}\left[\frac{k_{l,w}(T_{l,W} - T_{l,I})}{(\delta x)_w - (0.5 - f_P)(\Delta x)_P} - \frac{k_{l,e}(T_{l,I} - T_{l,E})}{(\delta x)_e + (0.5 - f_P)(\Delta x)_P} \right.$$
$$\left. + G_P(T_{e,P} - T_{l,P})(\Delta x)_P - C_{l,s}(T_{l,I})(T_{l,I} - T_{l,I}^0)\frac{(\Delta x)_P}{\Delta t}\right] \quad (65)$$

where $k_{l,w}$ and $k_{l,e}$ are lattice thermal conductivities at faces $w$ and $e$ of control volume, respectively, and $T_{l,I}^{0}$ is the solid-liquid interfacial temperature at the previous time step. The third and fourth terms in the bracket at the right-hand side of Eq. (65) represent the effects of electron-lattice interaction and change of the interfacial temperature on the solid-liquid interfacial temperature. Equations (65) will be used together with Eq. (58) to determine the solid-liquid phase interfacial velocity and temperature. The interfacial location at each step can be determined once the converged interfacial velocity is obtained.

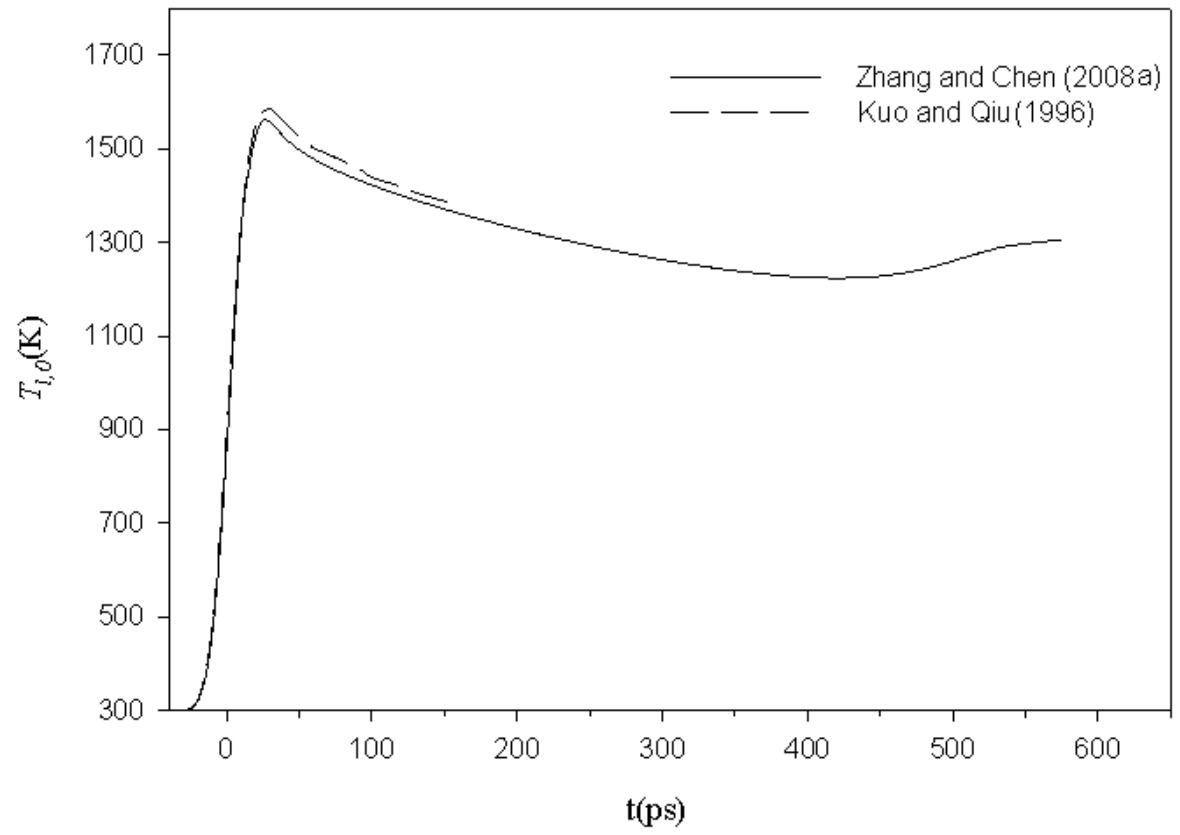

(a) Surface lattice temperature

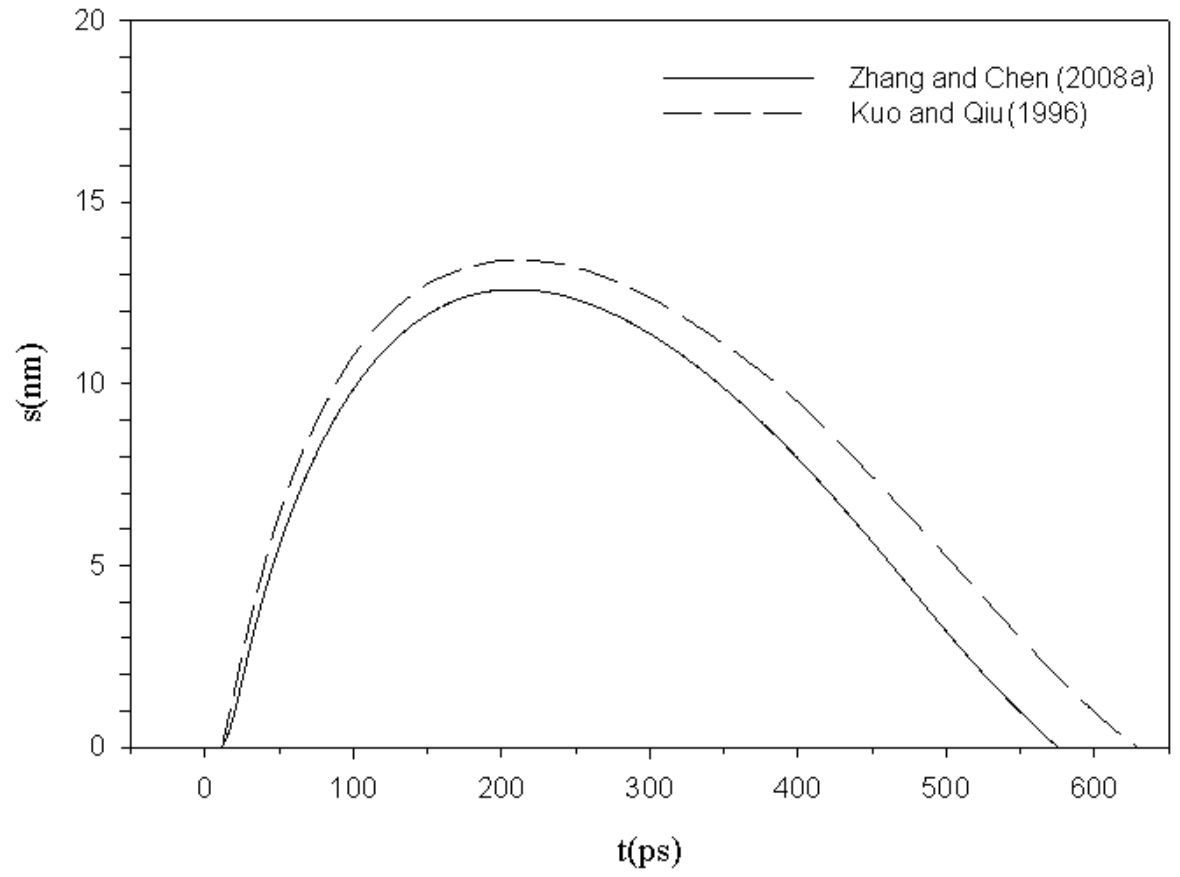

(b) Interfacial location

**Fig. 10** Comparison of surface lattice temperatures and interfacial locations (Zhang and Chen, 2008a)

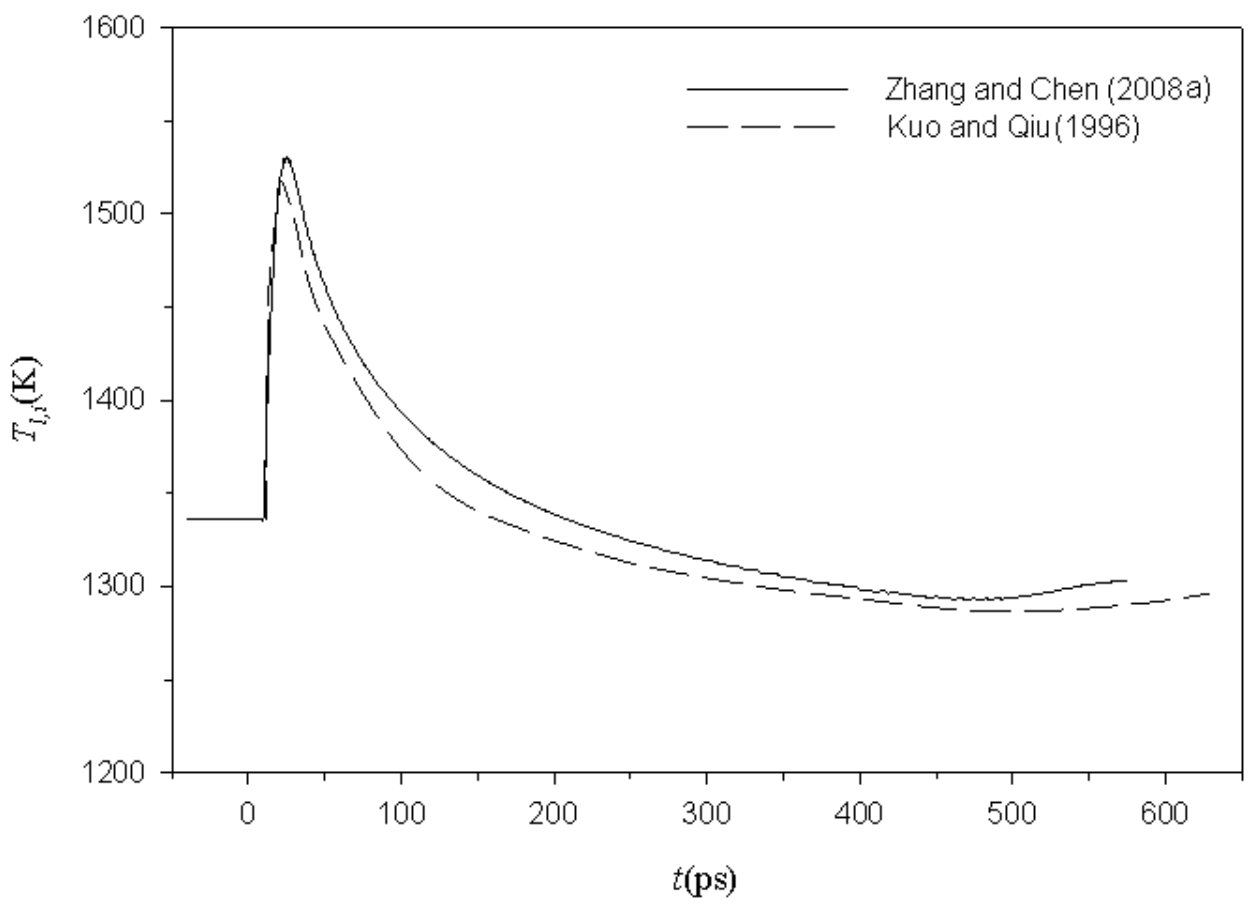

(a) Interfacial temperature

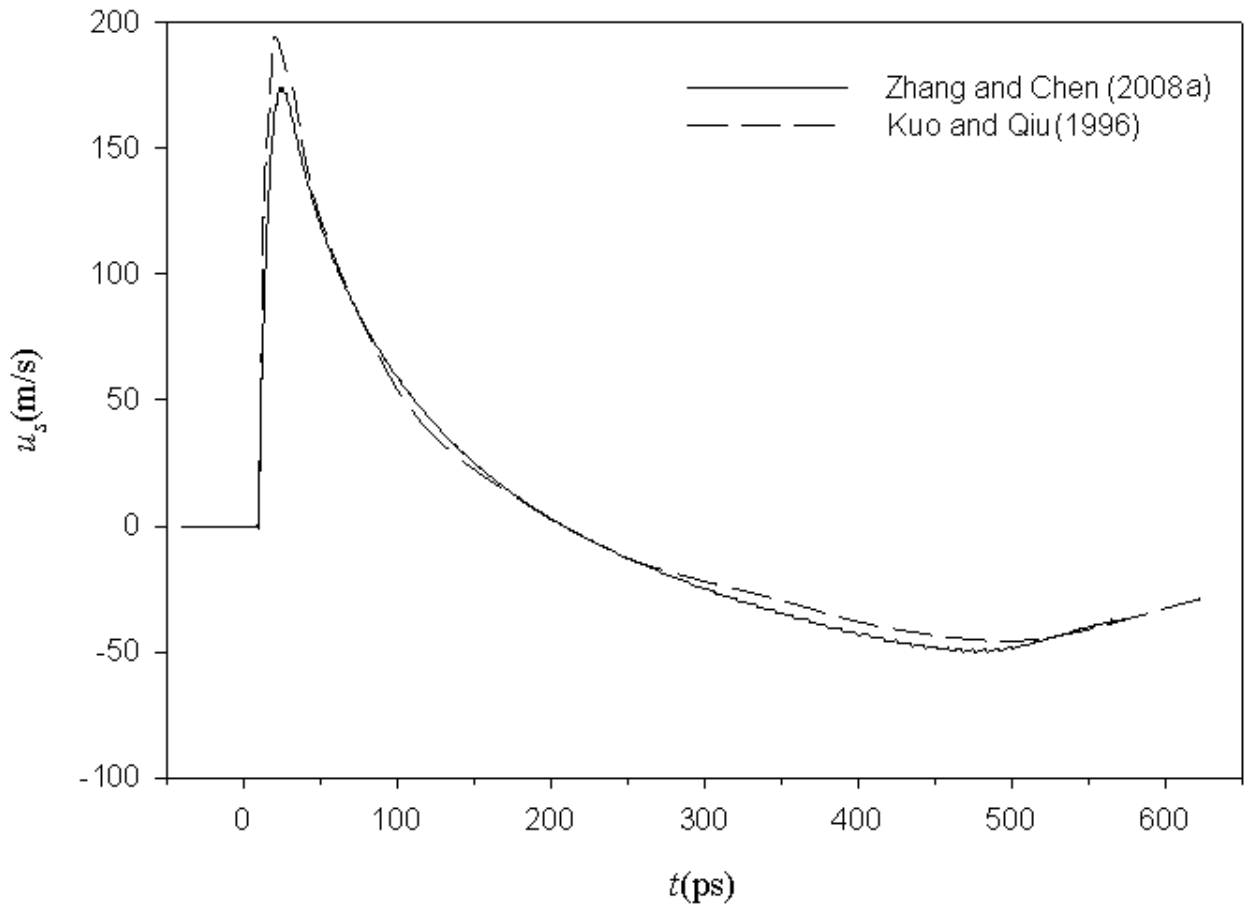

(b) Interfacial velocity

**Fig. 11** Comparison of interfacial temperatures and velocities (Zhang and Chen, 2008a)

Figure 10 shows comparison of the surface lattice temperature and interfacial location for a 1000-nm gold film under irradiation of a laser pulse with fluence of $J = 0.3$ J/cm$^2$ and a pulse width of $t_p = 20$ ps. The grid number is 2502 and the time step is $\Delta t = 0.05$ ps. The peak surface lattice temperature is 1562 K and occurs at $t = 27$ ps, while the maximum melting depth is 12.59 nm and occurs at $t = 209$ ps. The peak temperature and the maximum melting depth obtained in the present paper are 1.4% and 6% lower than those obtained by Kuo and Qiu (1996). The duration of phase change in the present paper is 9% shorter than that obtained by Kuo and Qiu (1996).

The interfacial temperature and velocity during the picosecond laser pulse and gold film interaction are shown in Fig. 11. Strong superheating in the melting stage and undercooling in the solidification stage can be observed in Fig. 11(a), although the degrees of superheating and undercooling in the present paper are slightly lower than that in Kuo and Qiu (1996). The peak interfacial velocity in the present paper is lower than that in Ref. Kuo and Qiu (1996), which is consistent with the result in Fig. 10(b). The interfacial velocity reached to its peak at very early stage of phase change ($t = 23.5$ ps), and the phase change continues until $t = 575$ ps. Since interfacial velocity decreases in most part of phase change process ($23.5\text{ ps} < t < 575\text{ ps}$), the melting depth obtained by the present implicit scheme is smaller than that obtained by explicit scheme.

### 4.2 Melting and Resolidification of Gold Film Irradiated by Nano- to Femtosecond Lasers

The thermal conductivity of the electrons given by Eq. (17) is valid when the electron temperature is much lower than the Fermi temperature, $T_F$, which is $6.42\times10^4$ K for gold. If the electron temperature is comparable to Fermi temperature, the thermal conductivity need to be calculated by Eq. (51). For high electron temperature, $\vartheta_e \gg 1$, Eq. (51) results in the well-known dependence $k_e \sim T_e^{5/2}$ that is the characteristics of low-density plasma. On the other hand, Eq. (51) will reduce to Eq. (17) under low electron temperature limit, $\vartheta_e \ll 1$. When the more sophisticated electron thermal conductivity model, Eq. (51), was employed, the interfacial velocity and temperature were significantly increased due to the strong deviation between the two thermal conductivity models at high temperatures (Zhang and Chen, 2008a).

While Eq. (56) is widely used in many existing works to describe the internal heat source induced by ultrafast laser, Wellershoff *et al.* (1999) and Hohlfeld *et al.* (2000) suggested that the ballistic motion and diffusion of the hot electrons spread the absorbed laser energy into much greater depth of electrons, especially for the s/p-band metals. This hot electron bath should be initially localized within either the ballistic range (for s/p-band metals) or the optical penetration depth. Therefore, it is necessary to incorporate the effect of the ballistic motion and hot electron diffusion by adding the ballistic range, $\delta_b$, to the optical penetration depth in Eq. (56), i.e.,

$$S = 0.94\frac{1-R}{t_p(\delta+\delta_b)[1-e^{-L/(\delta+\delta_b)}]}J\exp\left(\frac{x}{\delta+\delta_b}-2.77\left(\frac{t}{t_p}\right)^2\right) \qquad (66)$$

where $[1-e^{-L/(\delta+\delta_b)}]$ is to correct the film thickness effect. Consideration of the ballistic range of electron motion resulted in penetration of laser energy into deeper parts of the metal film and thus lowers interfacial velocity and temperature (Zhang and Chen 2008a).

The heat capacity of electron is assumed to be proportional to the electron temperature as it was indicated by Eq. (16). This assumption is valid for $T_e < 0.1T_F$ where $T_F$ is the Fermi temperature (Ashcroft and Mermin, 1976). When the electron temperature is greater than the Fermi temperature, the electron heat capacity can be obtained by ideal gas model, i.e., $C_e = 3Nk_B/2$, where $N$ is number density of atoms, and $k_B$ is Boltzmann constant. Jiang and Tsai (2005) obtained electron heat capacity by numerically integrating the relation

$$C_e(T_e) = n_e \left( \frac{\partial \langle \varepsilon \rangle}{\partial T_e} \right)_V \tag{67}$$

where $n_e$ is the number density of free electrons, $\langle \varepsilon \rangle$ is average kinetic energy per electron in J, and $V$ is the volume. The results showed that the heat capacity can be accurately represented by Eq. (16) for low temperature while ideal gas model is valid for higher temperature. Chen *et al.* (2006) analyzed the results of Jiang and Tsai (2005) and suggested that the electron heat capacity can be approximated by Eq. (43).

Most of the ultrafast laser heating analysis has been carried out with a constant electron-phonon coupling factor ($G$). Due to the significant changes in the electron and lattice temperature caused by high-power laser heating, $G$ could be temperature-dependent. The phenomenological temperature-dependent $G$ proposed by Chen *et al.* (2005) is adopted by Zhang and Chen (2007):

$$G = G_{RT}\left[ \frac{A_e}{B_l}(T_e + T_l) + 1 \right] \tag{68}$$

where $G_{RT}$ is the coupling factor at room temperature; $A_e$ and $B_l$ are materials constants for electron relaxation time. The coupling factor for liquid phase is enlarged by 20% from the result obtained from Eq. (68), because electrons collide more frequently with liquid atoms than with the atoms in the solid crystals.

Figure 12 shows comparison of interfacial locations, temperatures, and velocities obtained using different electron heat capacities and coupling factors. Since the highest electron temperature is less than $T_F / \pi^2$, the electron heat capacity obtained by Eqs. (16) and (43) are identical. The differences exhibited in Fig. 12 are attributed to different electron-phonon coupling factor only. It can be seen that the maximum melting depth is increased from 10.28 nm (occurring at $t = 199$ ps) for constant coupling factor to 17.30 nm (occurring at $t = 233.5$ ps) for temperature-dependent coupling factor. The maximum interfacial temperature in the melting stage is increased from 1499 K to 1581 K when the temperature-dependent coupling factor is considered. The peak interfacial velocity is increased from 151 m/s to 209 m/s due to the effect of temperature-dependent coupling factor. The results shown in Fig. 12 indicate that strong electron-lattice coupling results in higher lattice temperature, more rapid melting, and longer duration of phase change.

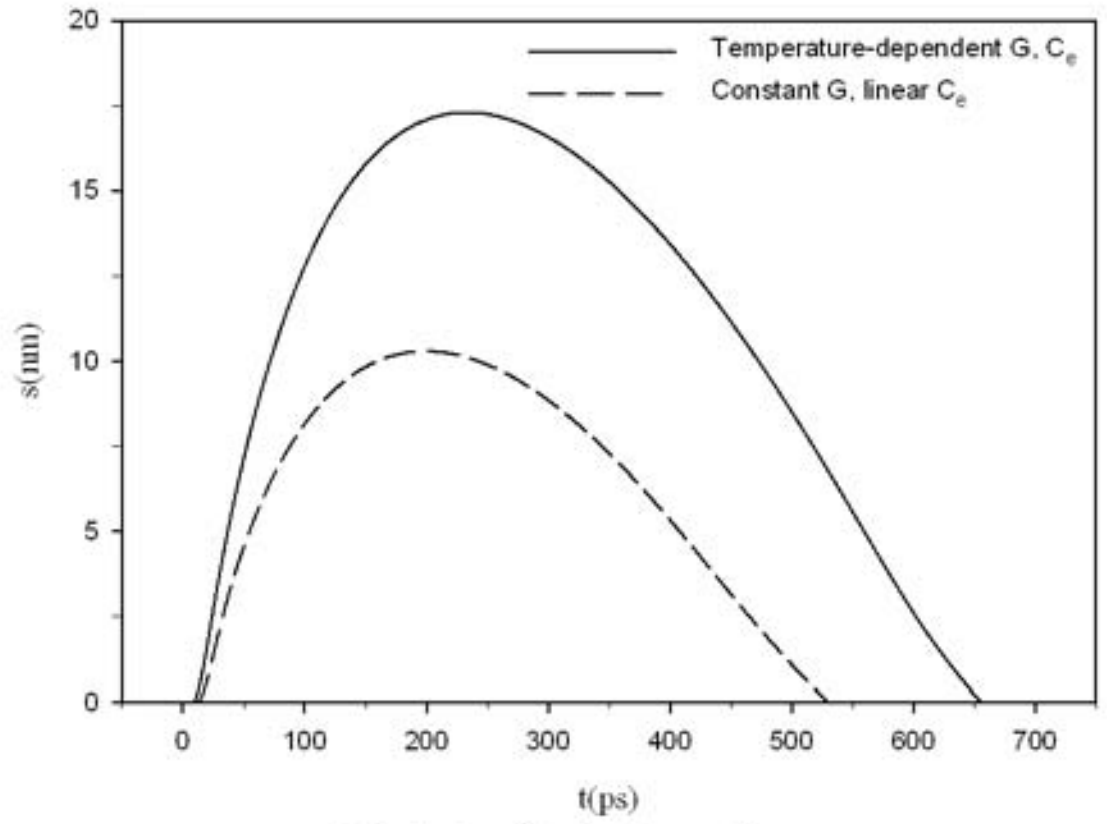

(a) Interfacial location

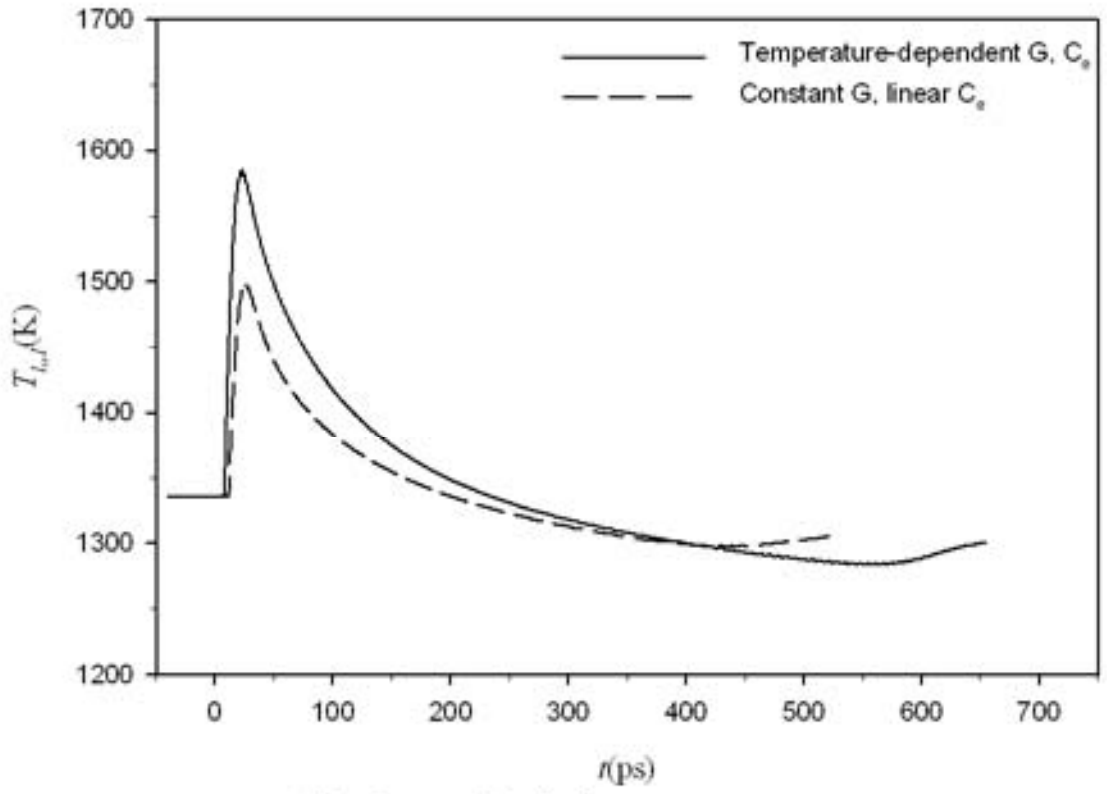

(b) Interfacial temperature

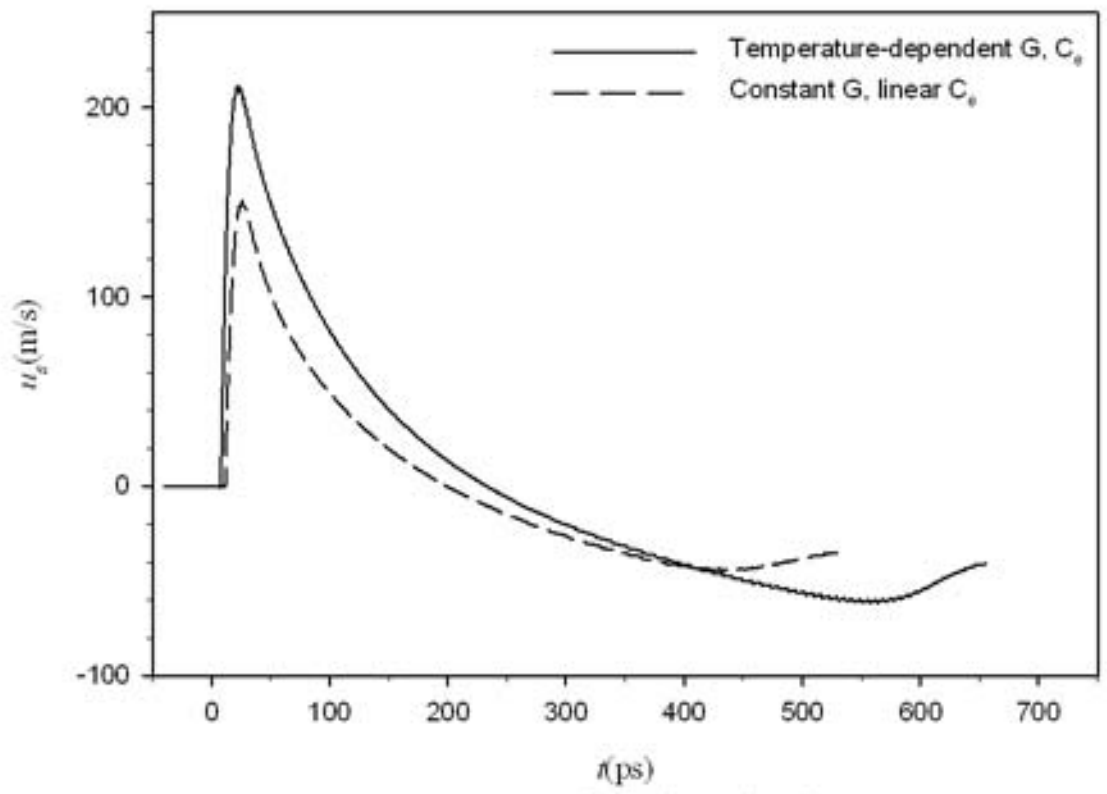

(c) Interfacial velocity

**Fig.** 12 Comparison of interfacial location, temperatures and velocities ($J$ = 0.3 J/cm$^2$, $t_p$ = 20 ps; Zhang and Chen, 2007)

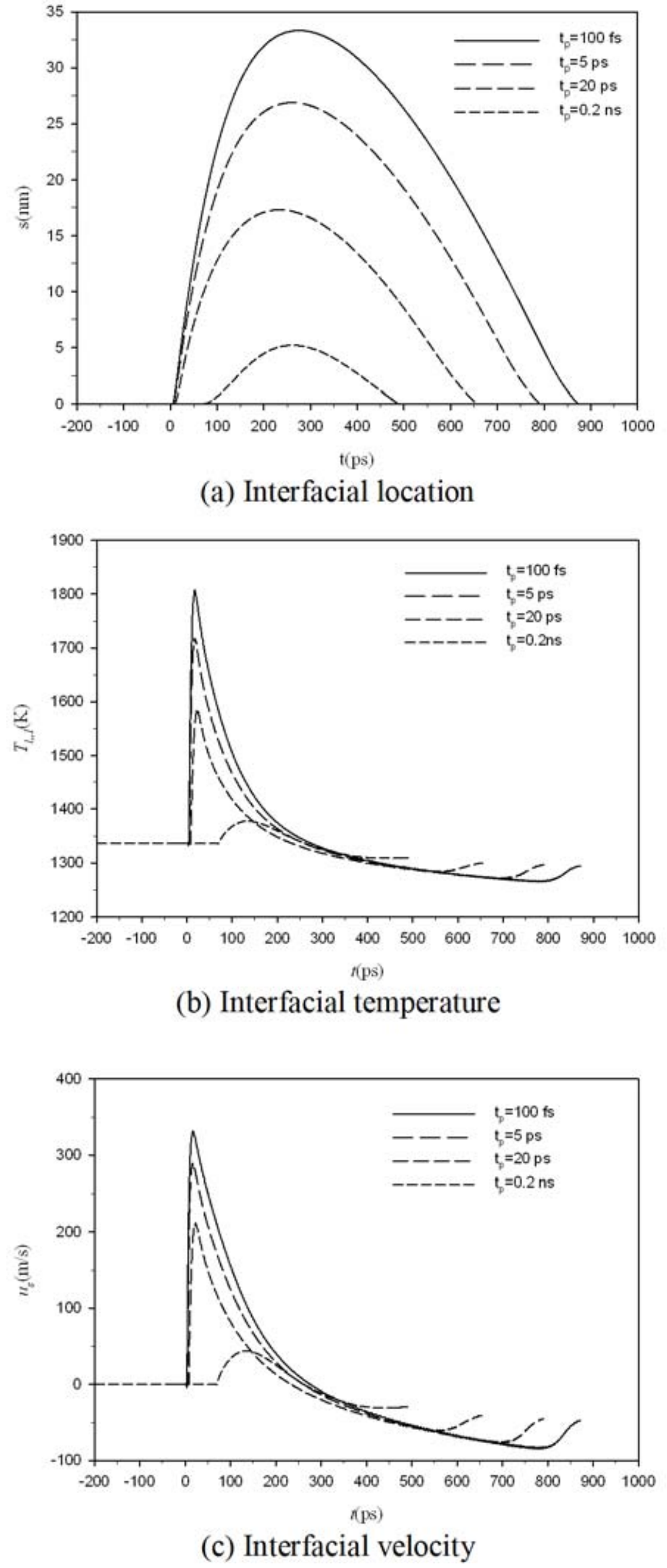

**Fig. 13** Effects of pulse width on the interfacial locations, temperatures, and velocities ($J$ = 0.3 J/cm$^2$; Zhang and Chen, 2007)

Figure 13 shows the effects of laser pulse width on the interfacial locations, temperatures, and velocities for a 1-μm gold film when the laser fluence is held at $J = 0.3$ J/cm$^2$. The grid number is 2502 and the time step varies from is 0.5 fs to 2 ps depending on the pulse widths. Although total laser energy delivered to the gold film are the same for all cases, concentration of laser energy into short pulse results in dramatic increase of the electron temperature. Since the electron temperature for $t_p = 5$ ps and $t_p = 100$ fs exceeded $T_F / \pi^2$, it is essential to use nonlinear electron heat capacity and temperature dependent coupling factor to model ultrafast melting and resolidification. It can be seen that both the maximum melting depth and duration of phase change increase with decreasing pulse width although the total laser energy delivered to the metal film are the same for all cases. The maximum melting depths for four different pulses width are 5.19 nm, 17.30 nm, 26.88 nm, and 33.32 nm, respectively. The maximum interfacial temperatures in the melting stage are 1378 K, 1585 K, 1718 K to 1808 K, respectively, and the corresponding peak interfacial velocities are 44 m/s, 211 m/s, 289 m/s and 332 m/s, respectively. Numerical simulation was also performed for laser pulse of $t_p = 1$ ns but no melting takes place because the peak surface lattice temperature was only 1080 K.

### 4.3 Microparticles

Selective Laser Sintering (SLS) is a rapid manufacturing/tooling technology that can build functional parts from powdered material via layer-by-layer sintering (for amorphous powder, such as polycarbonate) or melting (for crystalline powder, such as metal) by a directed laser beam (Beaman *et al.*, 1997). Fabrication of the metal part is very challenging because the temperature required to bind the metal powder particles is much higher than that to bind the amorphous powders particles. The single-component metallic parts can be fabricated by melting the skin of the powder particles and join together the non-melted solid cores by resolidifying the liquid layer (Tolochko *et al.*, 2003a,b). SLS of a single component metal powder via partial melting is very challenging due to the very small window of the processing parameters (Xiao and Zhang, 2006, 2007). Inappropriate combination of scanning velocity and laser beam intensity may result in non-melting or complete melting of metal powder particles.

It has been considered as a drawback of SLS technique that the fabricated metal parts are always porous and require post-processing to achieve fully densified (Das *et al.*, 1998). However, porosity is not always undesirable for some applications such as in wick structure of heat pipes, electrodes of fuel cells, aerospace, and bioengineering. Porous metallic materials, which are consistent of metal skeleton and beneficial pores, possess useful properties that the corresponding bulk materials do not have. Instead of avoiding the porosity in the product, it is possible to *utilize and control* the porosity in the final product by controlling the laser pulse width, repetition rate, laser intensity, and scanning velocity. While nanosecond laser can be used to sinter metal particles (Fischer *et al.*, 2002; Konrad, *et al.*, 2007), application of pico- to femtosecond lasers in sintering of metal powder particles may allow more accurate control of porosity of the sintered part via controlling the degree of partial melting. Because the pulse duration of femtosecond lasers is shorter than the relaxation times of most materials, and they have extremely high irradiance, the phenomena associated with femtosecond laser interaction with matter are very different from those associated with longer pulses. Sintering of metallic submicron or nanoparticles with femtosecond lasers has not been investigated.

Zhang and Chen (2008b) modeled ultrafast melting and resolidification of submicron particles irradiated by pico- to femtoseconds laser using the interfacial tracking method. The diameter of the metal particle is much smaller than the diameter of the laser beam, which is in turn much smaller than the dimension of the sintered part. Since the laser radiation penetrates the powder bed over a distance of several powder-sphere diameters, it can be assumed that multiple scattering of the radiation leads to a nearly homogeneous distribution of the heat flux within the optically penetrated layer (Fischer *et al.*, 2002), which leads to an almost normal incidence of the radiation on the surfaces of the grains in the underlying layers. Figure 14 shows the physical model of melting and resolidification under consideration. A particle with a radius of $r_o$ (diameter of $d_p$) and an initial temperature of $T_i$ is subjected to a temporal Gaussian laser pulse with a FWHM (Full Width at Half Maximum) pulse width of $t_p$ and fluence of $J$ ($J/m^2$) from the surface ($r = r_o$). Due to symmetry of the particle, the model can be simplified to be 1-D in the $r$-direction. The origin of time is chosen as the time when the laser pulse is at its maximum.

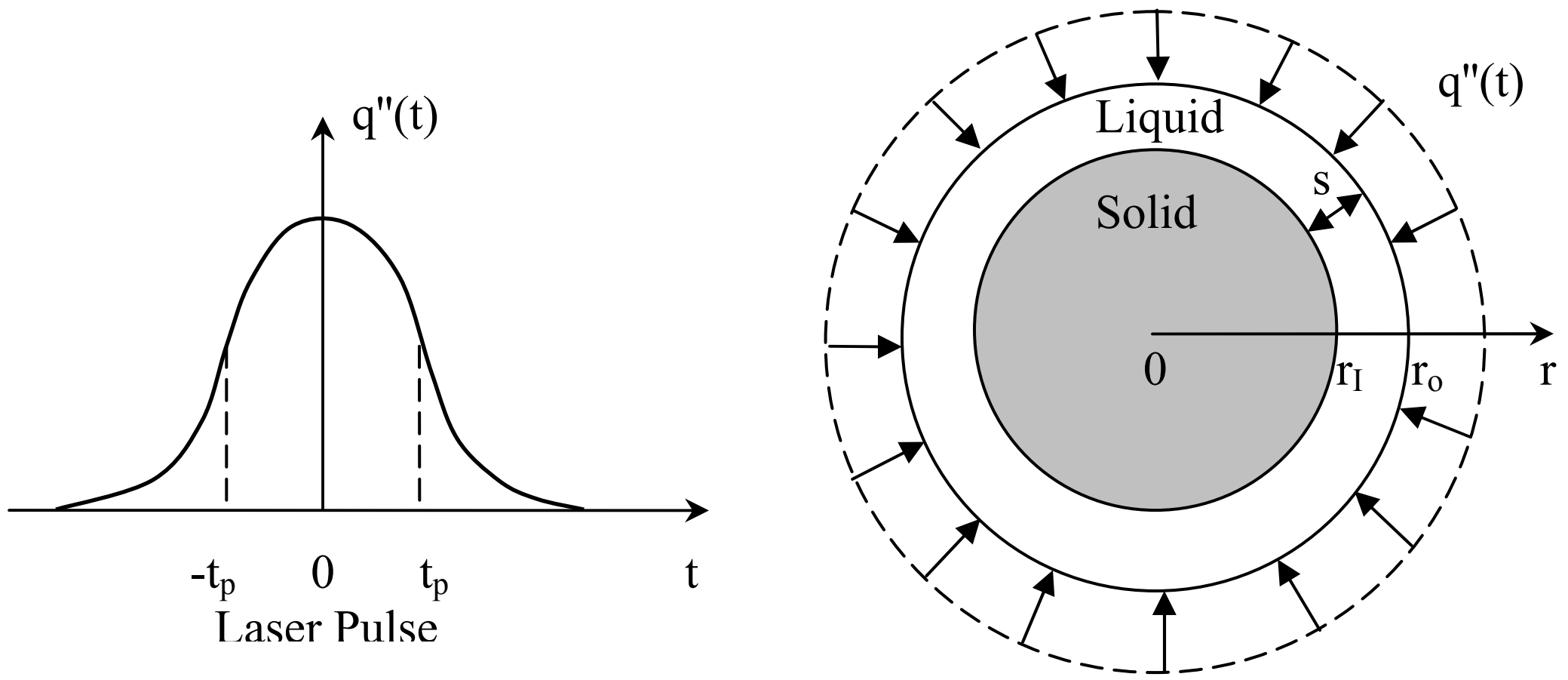

**Fig. 14** Ultrafast melting and resolidification of metal particle (Zhang and Chen, 2008b)

The energy equations of the free electrons ($e$) and lattice ($l$) can be obtained by simplifying Eqs. (14) and (25), i.e.,

$$C_e \frac{\partial T_e}{\partial t} = \frac{1}{r^2}\frac{\partial}{\partial r}\left[r^2 k_e \frac{\partial T_e}{\partial r}\right] - G\left(T_e - T_l\right) + S \tag{69}$$

$$C_l \frac{\partial T_l}{\partial t} = \frac{1}{r^2}\frac{\partial}{\partial r}\left[r^2 k_l \frac{\partial T_l}{\partial r}\right] + G\left(T_e - T_l\right) \tag{70}$$

The heat conduction in both electrons and lattice is assumed to be parabolic since the laser pulses considered are 100 fs or longer. The heat capacity of electron is obtained by Eq. (43). The thermal conductivity of the electrons, which depends on the temperatures of both electrons and lattice, is obtained form Eq. (51). The temperature-dependent electron-phonon coupling factor, $G$, was obtained from Eq. (68). The coupling factor for liquid phase is enlarged by 20% from the result obtained from Eq. (68), because electrons collide more frequently with liquid atoms than with the atoms in the solid crystals. The laser heat source term in Eq. (69) can be described by the following equation:

$$S = 0.94\frac{1-R}{t_p(\delta+\delta_b)[1-e^{-r_o/(\delta+\delta_b)}]}J\exp\left(\frac{r_o - r}{\delta+\delta_b} - 2.77\left(\frac{t}{t_p}\right)^2\right) \quad (71)$$

The time $t = 0$ is defined as the time when the peak of a laser pulse reaches the film surface. Therefore, the initial conditions of the problem are

$$T_e(x,-2t_p) = T_l(x,-2t_p) = T_i \quad (72)$$

The boundary conditions of the problem can be specified by assuming that the heat loss from the film surface can be neglected, i.e.,

$$\left.\frac{\partial T_e}{\partial r}\right|_{r=0} = \left.\frac{\partial T_e}{\partial r}\right|_{r=r_o} = \left.\frac{\partial T_l}{\partial r}\right|_{r=0} = \left.\frac{\partial T_l}{\partial r}\right|_{r=r_o} = 0 \quad (73)$$

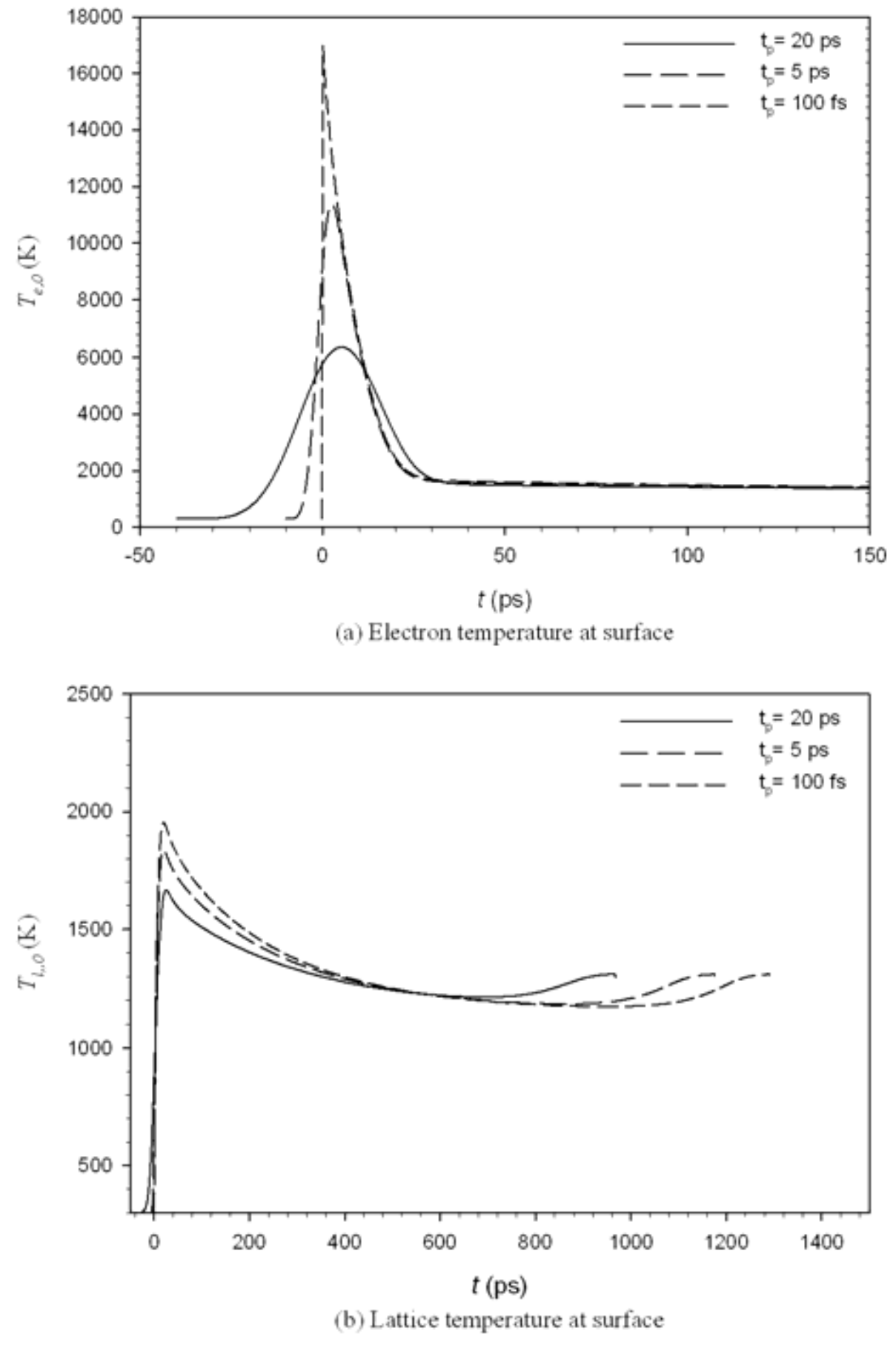

**Fig. 15** Effects of pulse width on surface electron and lattice temperatures ($J$ = 0.3 J/cm$^2$, $d_p$ = 1 μm; Zhang and Chen, 2008b)

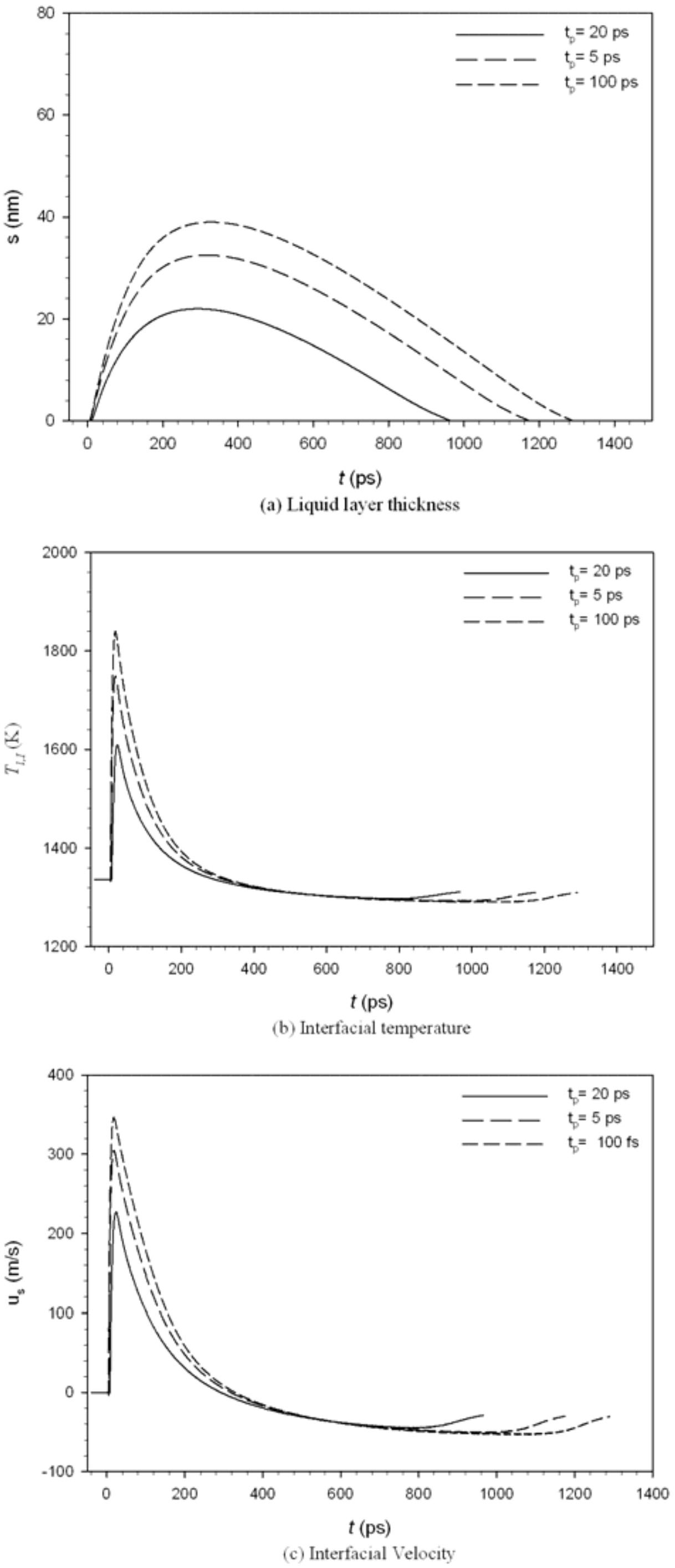

**Fig. 16** Effects of pulse width on the interfacial locations, temperatures, and velocities ($J = 0.3$ J/cm$^2$, $d_p = 1$ μm; Zhang and Chen, 2008b)

Figure 15 shows the effects of laser pulse width on the electron and lattice temperatures at the particle surface for a 1-μm gold particle when the laser fluence is held at $J = 0.3$ J/cm$^2$. The grid number is kept at 2502 while the time step varies from 1 fs to 2 ps depending on the pulse widths. For the three different pulse widths studied, the peak electron surface temperatures are 6,355 K, 11,391 K, and 16,983 K, respectively. With the same laser fluence of $J = 0.3$ J/cm$^2$, the total laser energy delivered to the gold particle are the same for all cases. However, concentration of laser energy into short pulse duration results in dramatic increase of the electron temperature. The peak lattice temperatures at the surface for the three different pulse widths are 1666 K, 1836 K, and 1953 K, respectively. It can be seen that the surface lattice temperature is much lower than the electron temperature, especially for shorter pulse width. As the pulse width decreases, there is longer delay from the time at which peak electron temperature occurs to the time at which peak lattice temperature occurs. Figure 16 shows the interfacial locations, temperatures, and velocities for the three pulse widths studied. With decreasing pulse width, both the maximum melting depth and duration of phase change increase. The maximum melting depths for the three pulse widths are 21.98 nm, 32.51 nm, and 38.99 nm, respectively. The maximum interfacial temperatures in the melting stage are 1609 K, 1749 K to 1836 K, respectively, and the corresponding peak interfacial velocities are 227 m/s, 304 m/s and 345 m/s. Thus, larger melting depth and higher interfacial velocity can be achieved by concentrating the same amount of laser energy into short pulse duration.

# 5. HOT-ELECTRON BLAST

## 5.1 Origin of the Electron Blast Force

The ways in which heat and load are transmitted through materials play a central role in understanding the refined mechanisms during ultrafast heating and deformation in the femtosecond domain. To inspire further interest in modeling the various competing mechanisms in the ultrafast processes, grooving on stainless steel by a nanosecond laser (a) and by a femtosecond laser (b) is shown in Fig. 17. Prolonged heating by nanosecond lasers causes melting, resulting in rough surfaces when the liquid phase reconsolidates back into the solid phase, as shown on the left. On the other hand, the same material processed by a femtosecond laser, as shown on the right, results in a strikingly smooth surface in the channel. The smooth surface may result from two different processes. In the first process, there is a spontaneous formation of bubbles in the melt during the femtosecond heating, which drastically increases from about 1 bubble/cm$^3$ to $10^{26}$ bubbles/cm$^3$ due to dynamic nucleation (Rethfield et al., 2002; Xu and Song 2000; Dyer et al., 2003). The melt is sputtered away as these bubbles rise altogether. The second process, the hot-electron blast (Tzou et al., 2005; Chen et al., 2006), results from the nonequilibrium heating of electrons in the femtosecond domain. Fast expansion of the electron gas results in a high pressure that is exerted on the surfaces of the metal lattices, resulting in separations of the lattices as the blasting force becomes excessive. In reality, femtosecond processes may be a combination of both, with the latter being a mechanical process that does not attribute to the phase change of the material.

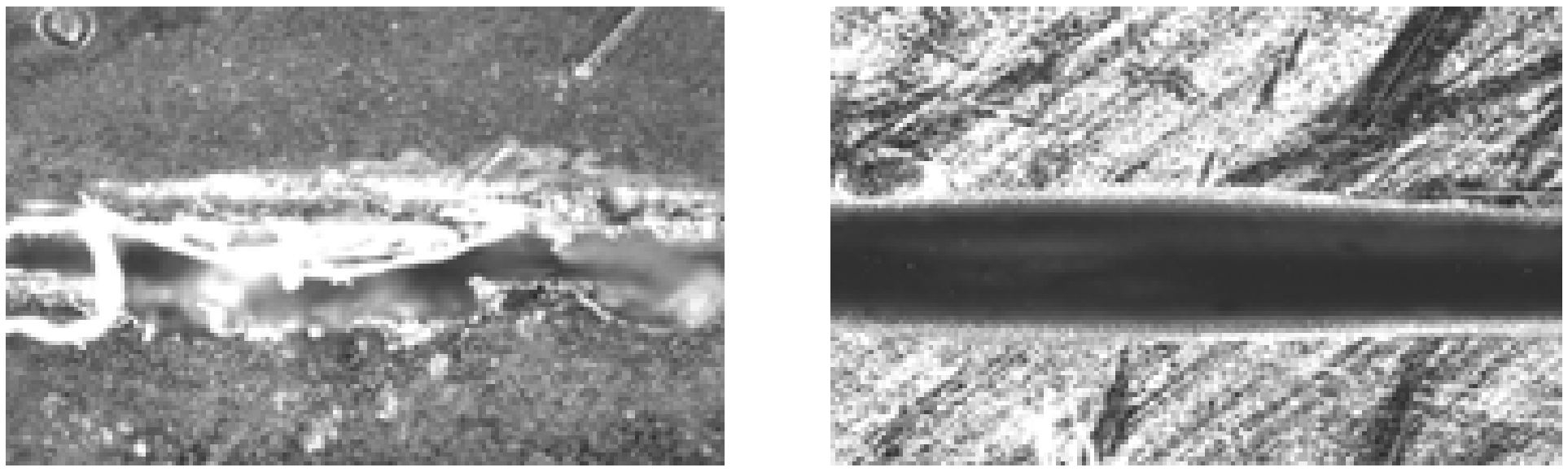

(a) Nanosecond laser (b) Femtosecond laser

**Fig. 17** Grooving on stainless steel by different lasers (Photos Courtesy of Lawrence Livermore Laboratory, the University of California)

The mechanical force generated by femtosecond lasers is counterintuitive in nature. Unlike the continuous or long-pulse laser heating where damage by the thermal process prevails, femtosecond lasers can produce high quality surfaces with minimal collateral damage by the ways the mechanical forces are produced and their intrinsic effects on the ultrafast deformation of the lattices. With the fundamental understanding developed in the previous works (Tzou et al., 2005) during the sub-picosecond transient, this work reinstates the effects of volumetric expansion/contraction of the metal lattices, thermal wave effect for heat propagation in phonons, and thermomechanical coupling through the temperature gradient across the metal lattices. These physical mechanisms follow the hot-electron blast developed in the femtosecond domain, becoming highly activated shortly after the laser pulse. The thermomechanical model incorporates all these effects is therefore able to describe the ultrafast thermomechanical response covering approximately three orders of magnitude, from tens of femtoseconds (where electron-to-phonon interactions dominate) to tens of picoseconds (where phonon-to-phonon interactions take over). While accommodating the temperature dependent thermal properties of electrons, the thermophysical parameters that characterize the ultrafast transient are extracted in terms of nondimensional groups. Their values shall be varied over one to two orders of magnitude in studying their effects on the ultrafast heating and lattice deformation. The finite-difference differential formulation shall be developed to tackle the nonlinearly coupled partial differential equations (PDEs), and the formulation shall be extended to study the effect of repetitive pulsing, as well as the potential effect of plasma shielding due to the surface ionization induced by the UUL.

### 5.2 Ultrafast Deformation – Electron Blast

Due to the significant deviation of the lattice temperature from the reference temperature, particularly beyond 10 picoseconds (threshold value of the relaxation time for phonons) where the metal lattices become highly heated by the hot electrons, the linear theory of thermoelasticity breaks down and nonlinear coupling between the thermal and mechanical fields become intrinsic in ultrafast thermomechanics. With all these effects interwoven during the ultrafast transient, the field equations governing the ultrafast response are extremely complicated and nonlinear.

Ultrafast heating is highly nonequilibrium, necessitating separate considerations of energy transport in electrons and phonons with the electron-phonon thermal coupling. The energy equations for electrons and lattices are:

$$C_e \frac{\partial T_e}{\partial t} = \nabla \cdot \left(k_e \nabla T_e\right) - G\left(T_e - T_l\right) + S \quad (74)$$

$$C_l \frac{\partial T_l}{\partial t} + \left(\frac{C_l \eta}{\kappa_\varepsilon}\right) \frac{\partial}{\partial t} (\nabla \cdot \mathrm{u}) = -\nabla \cdot \mathrm{q}_l'' + G\left(T_e - T_l\right) \quad (75)$$

where the heat flux for the lattice, $\mathbf{q}_l''$, satisfies Eq. (27), $\tau$ the phonon-to-phonon relaxation time, $\eta$ the thermomechanical coupling factor, and $\kappa_\varepsilon$ the thermal expansion coefficient of strain. Thermal relaxation of electrons is neglected since the electron-to-electron relation time is only for a few femtoseconds. On the other hand, consumption of thermal energy in support of the thermal expansion/contraction of the lattice volume has now been included in Eq. (75) due to the high time-rate of change of the lattice volume $\partial(\nabla \cdot \mathrm{u})/\partial t$ in the femtosecond domain. This effect is important because during the fast transient in the femtosecond domain, a volumetric strain $\nabla \cdot \mathrm{u}$ of the order of nanometers would result in a strain rate $\partial(\nabla \cdot \mathrm{u})/\partial t$ on the order of $10^6$. Under constant phonon properties, the heat flux vector $q_l''$ in phonons can be eliminated from Eq. (75) by considering Eq. (27), resulting in

$$k_l \nabla^2 T = C_l \left(\frac{\partial T_l}{\partial t} + \tau \frac{\partial^2 T_l}{\partial t^2}\right) - G\left[\left(T_e - T_l\right) + \tau\left(\frac{\partial T_e}{\partial t} - \frac{\partial T_l}{\partial t}\right)\right] + \frac{C_l \eta}{\kappa_\varepsilon}\left[\frac{\partial}{\partial t}\left(\nabla \cdot \mathrm{u}\right) + \tau \frac{\partial^2}{\partial t^2}\left(\nabla \cdot \mathrm{u}\right)\right] \quad (76)$$

A number of high-order derivatives result due to the high-rate responses, including the lagging behavior describing the delayed response between the heat flux vector and the temperature gradient in heat transport (Tzou, 1997). Consequently, special methods developed in classical thermoelasticity assuming Fourier's law heat diffusion, such as the Goodier's potential or Boussinesq-Papkovich functions (Boley and Weiner, 1960), no longer hold.

Electrons can be viewed as a special type of gas surrounding the metal lattices (phonons) (Tzou et al., 2005; Chen et al., 2006). When highly excited, the electron gas rapidly expands, which could result in excessive pressure being exerted on the surfaces of the metal lattices. The equations describing the motion of the metal lattices can be written as

$$\rho \frac{\partial^2 \mathbf{u}}{\partial t^2} = \nabla \cdot \boldsymbol{\sigma} + \nabla P_e \quad (77)$$

where $\boldsymbol{\sigma}$ is Cauchy stress tensor expressed as

$$\boldsymbol{\sigma} = \lambda(\nabla \cdot \mathbf{u})\mathbf{I} + \mu\left[(\nabla \mathbf{u}) + (\nabla \mathbf{u})^T\right] - \kappa_\sigma\left(T_l - T_0\right)\mathbf{I} \quad (78)$$

where $\rho$ is the mass density, $\mathbf{u}$ is the displacement vector of the metal lattice, $\mathbf{I}$ is unit tensor, $\lambda$ and $\mu$ are the Lamé constants, $\kappa_\sigma$ is the thermal expansion coefficient of stress. Kinetic pressure of the electron gas is represented by:

$$P_e = nk_B T_e \quad \text{with} \quad n = \frac{\pi^2 N}{3}\left(\frac{T_e}{T_F}\right) \quad \Rightarrow \quad P_e = \Lambda T_e^2 \quad \text{with} \quad \Lambda = \frac{\pi^2 N k_B}{3T_F} \tag{79}$$

with $n$ representing the number density of electrons proportional to the electron temperature ($T_e$) and number density of atoms ($N$) and $k_B$ representing the Boltzmann constant. Neglecting the quantum effect, Eq. (79) can be viewed as the equation of state for the electron gas, where the kinetic pressure of electrons is proportional to the electron temperature squared. Substituting Eq. (79) into Eq. (77) results in

$$\rho \frac{\partial^2 \mathbf{u}}{\partial t^2} = \nabla \cdot \boldsymbol{\sigma} + 2\Lambda T_e \nabla T_e, \tag{80}$$

which displays a driving force due to the rapid expansion of hot electrons. Such a blasting force exerting on the lattice surfaces is proportional to both the temperature and temperature gradient of the electron gas, which will be significant in the early stage of electron heating as described by Eq. (74). Combination of the Eqs. (77) and (78) gives

$$\rho \frac{\partial^2 \mathbf{u}}{\partial t^2} = (\lambda + \mu)\nabla\left(\nabla \cdot \mathbf{u}\right) + \mu \nabla^2 \mathbf{u} + \Lambda T_e \nabla T_e - \kappa_\sigma \nabla T_l \tag{81}$$

which includes explicit effects from both $T_e$ (through the hot-electron blast) and $T_l$ (thermomechanical coupling) on the motion of the metal lattices.

Equations (74), (76), and (81) were derived from the Boltzmann transport equation (Qiu and Tien, 1993) and the nonequilibrium partition functions for electrons (Falkovsky and Mishchenko, 1999). Rather than solving them in the microstructural frameworks for electrons and phonons, the continuum formulation as shown absorbs the microstructural effects in additional terms led by the new coefficients. They are the relaxation time ($\tau$) and electron-phonon coupling factor ($G$) in Eq. (76) and the electron pressure constant ($\Lambda$) in Eq. (81). This formulation should hold as long as the physical ground for stress (and consequently the elastic moduli in general) and temperature (hence the thermal conductivity or diffusivity) holds, which however will breakdown as the conductor is composed of only a few lattices.

### 5.3 One-Dimensional Example

To demonstrate the capability of the above thermomechanical model, Tzou and Pfautsch (2008) presented a finite difference solution of a one-dimensional example. Unique features in ultrafast heating and deformation, as shown by Equations (74), (76), and (81), lie in the thermomechanical response in time. Consideration of a one-dimensional case, in space, therefore, is sufficient to develop the full resolution during the ultrafast transient. Considering the lattice motion in a one-dimensional film with lateral constraints, the plane strain condition applies and the one-dimensional forms of Eq. (74), (76), and (81) are

$$C_e T_e \frac{\partial T_e}{\partial t} = \frac{\partial}{\partial x}\left(k_e \frac{\partial T_e}{\partial x}\right) - G\left(T_e - T_l\right) + S \tag{82}$$

$$k_l \frac{\partial^2 T_l}{\partial x^2} = C_l \left(\frac{\partial T_l}{\partial t} + \tau \frac{\partial^2 T_l}{\partial t^2}\right) - G\left[\left(T_e - T_l\right) + \tau\left(\frac{\partial T_e}{\partial t} - \frac{\partial T_l}{\partial t}\right)\right]$$

$$+\frac{C_l\eta}{\kappa_\varepsilon}\left(\frac{\partial^2 u}{\partial t\partial x}+\tau\frac{\partial^3 u}{\partial t^2\partial x}\right) \quad (83)$$

$$\rho\frac{\partial^2 u}{\partial t^2}-E_1\frac{\partial^2 u}{\partial x^2}+E_2\kappa_\sigma\frac{\partial T_l}{\partial x}=2\Lambda\left(T_e\frac{\partial T_e}{\partial x}\right) \quad (84)$$

where

$$E_1=\frac{E(1-\nu)}{(1+\nu)(1-2\nu)},\quad E_2=\frac{E}{1-2\nu}. \quad (85)$$

The spatial variable $x$ in Eqs. (82) – (84) describes the distance measured from $x$ = 0, the surface subject to the laser irradiation. The heat capacity and thermal conductivity of the electrons gas are obtained from Eq. (16) and (17), respectively. The internal heat source in Eq. (82) is obtained by Eq. (56). Heat loss from the film surfaces is negligible during the femtosecond transient,

$$\frac{\partial T_e}{\partial x}=0,\quad q_l''=0 \quad \text{at} \quad x=0 \quad \text{and} \quad x=l \quad (86)$$

with $L$ denoting the film thickness. The front and rear surface of the film is assumed strain-free,

$$\frac{\partial u}{\partial x}=0 \quad \text{at} \quad x=0 \quad \text{and} \quad x=l, \quad (87)$$

whereas the film is assumed to be heated from a stationary state:

$$T_e=T_0,\quad T_l=T_0,\quad \frac{\partial T_l}{\partial t}=0,\quad \text{and} \quad \frac{\partial u}{\partial t}=0 \quad \text{as} \quad t=t_0=-2t_p \quad (88)$$

The strain-free boundary conditions described in Eq. (87) are employed to tackle the ultrafast response in a simpler situation. For stress-free conditions, Eq. (87) is replaced by

$$\sigma=E_1\frac{\partial u}{\partial x}-E_2\kappa_\varepsilon T_l=0,\quad \text{or} \quad \frac{\partial u}{\partial x}=\left(\frac{E_2}{E_1}\right)\kappa_\varepsilon T_l \quad \text{at} \quad x=0 \quad \text{and} \quad x=l \quad (89)$$

which relate the surface strains to the surface temperatures of phonons.

Defining the following dimensionless variables and parameters:

$$\xi=\frac{x}{\delta},\quad \beta=\frac{t}{t_p},\quad \theta_{e(l)}=\frac{T_{e(l)}}{T_0},\quad U=\frac{u}{\left(\Lambda T_0^2 t_p^2/\rho\delta\right)},\quad Y=\frac{E_0/\rho}{(\delta/t_p)^2},\quad \beta_0=\frac{t_0}{t_p},$$

$$S=0.94J\left(\frac{1-R}{\delta C_{e0}T_0}\right),\quad H=\frac{Gt_p}{C_{e0}},\quad K=\frac{\alpha_0 t_p}{\delta^2},\quad C=\frac{C_{e0}}{C_l},\quad L=\frac{l}{\delta},\quad \alpha_0=\frac{k_{e0}}{C_{e0}}$$

$$M=\frac{\eta}{\left(\frac{\rho\delta^2\kappa_\varepsilon}{\Lambda T_0 t_p^2}\right)},\quad \Gamma=\frac{\tau}{t_p},\quad Y_1=\frac{E_1 t_p^2}{\rho\delta^2},\quad Y_2=\frac{E_2\kappa_\varepsilon}{\Lambda T_0}, \quad (90)$$

Eqs. (82) – (84) become

$$\theta_e\frac{\partial\theta_e}{\partial\beta}=K\frac{\partial}{\partial\xi}\left(\frac{\theta_e}{\theta_l}\frac{\partial\theta_e}{\partial\xi}\right)-H\left(\theta_e-\theta_l\right)+S\exp\left[-\xi-4\ln(2)\beta^2\right] \quad (91)$$

$$KC\frac{\partial^2\theta_l}{\partial\xi^2}=\Gamma\frac{\partial^2\theta_l}{\partial\beta^2}+\frac{\partial\theta_l}{\partial\beta}-HC\left[\left(\theta_e-\theta_l\right)+\Gamma\frac{\partial}{\partial\beta}\left(\theta_e-\theta_l\right)\right]$$
$$+M\left(\frac{\partial^2 U}{\partial\beta\partial\xi}+\Gamma\frac{\partial^3 U}{\partial\beta^2\partial\xi}\right) \tag{92}$$
$$\frac{\partial^2 U}{\partial\beta^2}-Y_1\frac{\partial^2 U}{\partial\xi^2}+Y_2\frac{\partial\theta_l}{\partial\xi}=2\theta_e\frac{\partial\theta_e}{\partial\xi} \tag{93}$$

Ultrafast heating and deformation is thus characterized by 7 parameters: $K$ (nondimensional thermal diffusivity of electrons), $H$ (nondimensional electron-phonon coupling factor), $S$ (nondimensional laser fluence), $C$ (heat-capacity ratio), $\Gamma$ (nondimensional relaxation time of phonons), $M$ (nondimensional group of thermomechanical coupling), $Y_1$ (nondimensional elastic modulus), and $Y_2$ (nondimensional thermoelastic modulus). Equations (91) – (93) provide three equations to be solved for three unknowns, $\theta_e$, $\theta_l$, and $U$, subjected to the initial and boundary conditions:

$$\theta_e=\theta_l=1,\quad \frac{\partial\theta_l}{\partial\beta}=0,\quad \text{and}\quad \frac{\partial U}{\partial\beta}=0\quad \text{as}\quad \beta=-2; \tag{94}$$
$$\frac{\partial\theta_e}{\partial\xi}=0,\quad \frac{\partial\theta_l}{\partial\xi}=0,\quad \frac{\partial U}{\partial\xi}=0\quad \text{at}\quad \xi=0\quad \text{and}\quad x=L. \tag{95}$$

Numerical solutions are obtained for a gold film with the following thermal and optical properties: $k_{e0}$ = 315 W/mK, $C_{e0} = 2.1 \times 10^4$ J/m$^3$K, $\tau$ = 10 ps, $C_l = 2.5 \times 10^6$ J/m$^3$K, $g = 2.6 \times 10^{16}$ J/m$^3$K, $\delta$ = 15.3 nm, $R$ = 0.93, $T_0$ = 300 K, and $l$ = 0.02 μm (Tzou, 1997). Mechanical properties for gold are $E$ = 79 GPa, $\nu$ = 0.42, $\kappa_\varepsilon = 14.2 \times 10^{-6}$, and $\Lambda = 10^5$ J/m$^3$K. The laser fluence is taken to be $J$ = 732 J/m$^2$. The nondimensional parameters are thus $H \cong 0.124$, $K \cong 6.41$, $C \cong 8.4 \times 10^{-3}$, $\Gamma \cong 100$, $M = 6.4915 \times 10^{-5}$, $Y_1 = 4.46376 \times 10^{-4}$, $Y_2 = 0.27156$, $S_0 = 500$, and $L$ = 1.307. These values shall be used throughout the numerical analyses unless stated otherwise.

Figure 18 describes the straining patterns developed in the gold film as the parameter $H$ varies from 0 (no energy exchange between electrons and phonons) to 12.4. The hot electron blast, $2\theta_e(\partial\theta_e/\partial\xi)$ in Eq. (93), introduces compressive strains ($U' < 0$) on the initial contact ($\beta = 0$) with the film (near $\xi = 0$), as indicated in Fig. 18(a). The peak of the compressive strain is located between the characteristic lines of thermal and mechanical waves (Tzou et al., 2005). After the initial compressive strain near the front surface, two tensile ripples follow, in the direction of increasing time ($\beta$ increases) or in the direction toward the interior of the film ($\xi$ increases). Another compressive ripple is resulted near the back surface of the film ($\xi > 0.8$) at longer times ($\beta > 20$). Such alterations between compressive and tensile strains prevail in the temporal and spatial directions, which become even more exaggerated at longer times due to reflections of the displacement waves from the front ($\xi = 0$) and rear ($\xi = 1.307$) surfaces of the film. As the value of $H$ increases, the peak values of compressive and tensile strains slightly increase. All qualitative features of the straining patterns, however, remain the same.

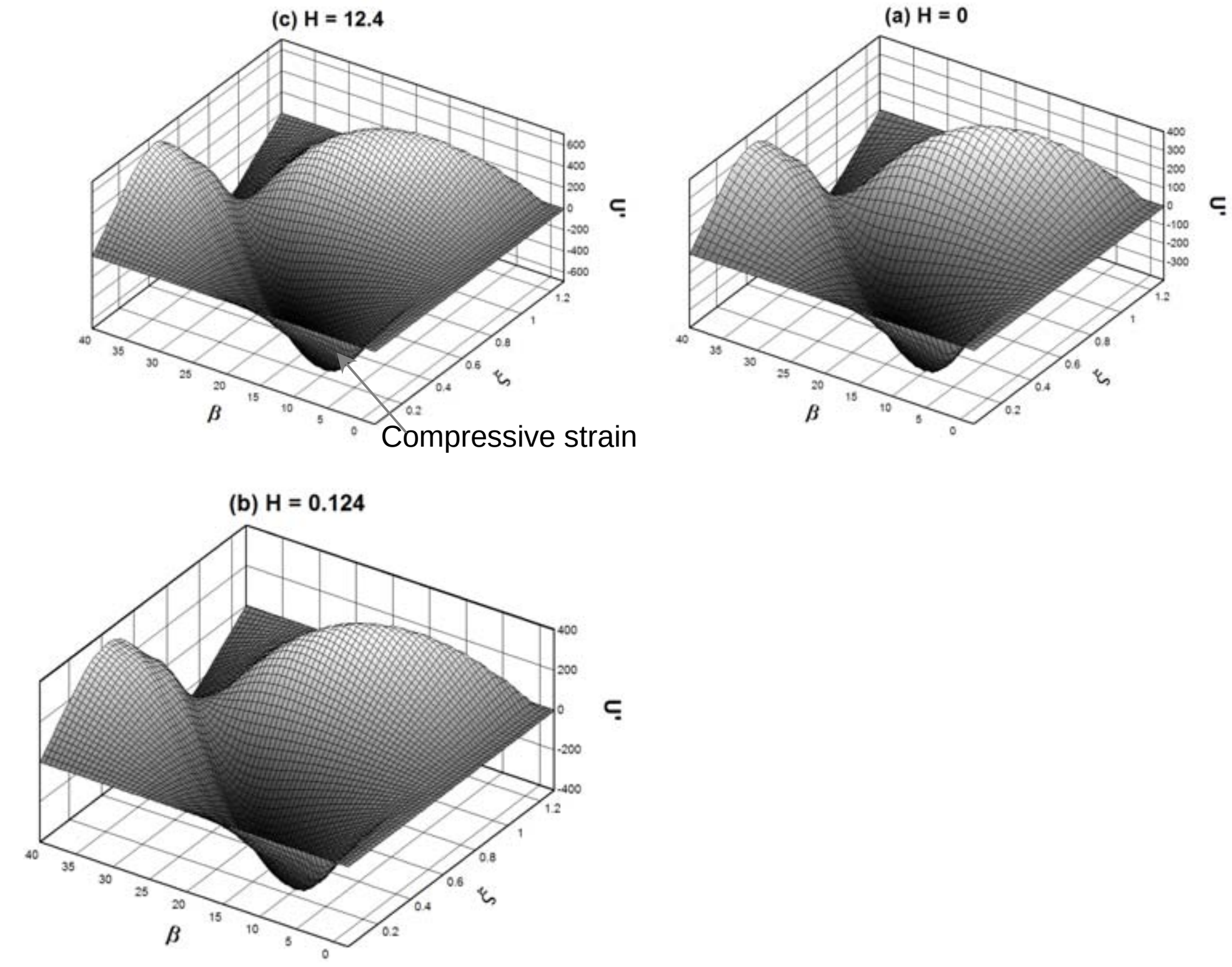

**Fig. 18** Straining patterns developed in the gold film (Tzou and Pfautsch, 2008)

# 6. CONCLUSION

Micro- and nanoscale heat transfer in femtosecond laser processing of metals is reviewed in this chapter. The classical heat conduction theory based on Fourier's law assumes that thermal disturbance propagates with an infinite speed. As heat conduction is accomplished by successive collision of the energy carriers (phonons or electrons), the prorogation of thermal disturbance is always at a finite speed. To account for the finite propagation speed of thermal wave, the Cattaneo-Vernotte thermal wave mode can be used. While the thermal wave model assumes that the temperature gradient always precedes the heat flux, the dual-phase lag model allows either the temperature gradient (cause) to precede heat flux vector (effect) or the heat flux vector (cause) to precede the temperature gradient (effect). The nonequilibrium between electrons and lattice has been observed experimentally and can be described by the classic two-temperature model. Various two-step heating models can be employed to describe such nonequilibrium phenomena between the electrons and lattices.

A semi-classical two-step heating model to investigate thermal transport in metals caused by ultrashort laser heating is also discussed. The semiclassical two-step heating model includes the conservation equations of number density, momentum and energy for electron subsystem. The main difference between the semiclassical and the phenomenological two-temperature models is that the former includes the effects of electron drifting, which could result in significantly different electron and lattice temperature response from the latter for higher-intensity and shorter-pulse laser heating. The volumetric laser heat source used in the model was modified for accommodating the electron ballistic motion and the delayed rise time of the electron temperature. The temperature fields in the electrons and lattice were solved with the equations of the conservation of momentum and energy in electron subsystem and the equation of heat conduction in the lattice subsystem. It was shown that for higher-intensity and shorter-laser pulses, the semiclassical two-temperature model could result in different thermal response than the phenomenal two-temperature model.

Under higher laser fluence and/or short pulse, the lattice temperature can exceed the melting point and melting takes place. The liquid phase will be resolidified when the lattice is cooled by conducting heat away. A fixed grid interfacial tracking method is presented to solve kinetics controlled rapid melting and resolidification during ultrashort pulse laser interaction with a free-standing metal film. Melting and resolidification were modeled by considering the interfacial energy balance and nucleation dynamics. An iterative solution procedure was employed to determine the elevated melting temperature and depressed solidification temperature in the ultrafast phase-change processes. Nonlinear electron heat capacity and temperature-dependent electron-lattice coupling factor on the rapid phase change are also investigated. Ultrafast melting and resolidification of a submicron gold particle subject to pico- to femtosecond laser pulse are studied.

At even shorter pulse width, femtosecond laser heating on metals produces a blasting force from hot electrons in the sub-picosecond domain, which exerts on the metal lattices along with the non-equilibrium heat flow. Physical phenomena of the hot-electron blast induced by ultrafast, ultraintense lasers have been further explored to include the finite speed of heat propagation in phonons, energy consumption in support of the volumetric expansion/contraction of the metal lattices, thermomechanical coupling, and thermal relaxation of phonons. Nonequilibrium heating emphasizes different temperatures of phonons and electrons before they come to thermal equilibrium. Inertia effect during the lattice motion and effect of volumetric expansion rate that have often been neglected in classical thermoelasticity have been found important in ultrafast deformation.

## ACKNOWLEDGEMENT

The work presented in this chapter is supported in part by the US National Science Foundation under grant number CBET -0730143.

## NOMENCLATURE

| | |
|---|---|
| $A_e$ | materials constants for electron relaxation time [1/K$^2$-s] |
| $B_l$ | materials constants for electron relaxation time [1/K-s] |
| $B_e$ | Coefficient for electron heat capacity [J/m$^3$-K$^2$] |
| $c$ | thermal propagation speed [m/s] |
| $C$ | heat capacity, [J/m$^3$-K] |
| $c_p$ | specific heat, [ J/kg-K ] |
| $E$ | Young's modulus [Pa] |
| $f$ | distribution function of the free electrons<br>liquid fraction in control volume |
| $G$ | electron-lattice coupling factor [ $W/m^3$-K ] |
| $H$ | nondimensional electron-phonon coupling factor |
| $H_l$ | enthalpy of lattice per unit volume [ $J/m^3$ ] |
| $h_m$ | latent heat of fusion [ J/kg ] |
| $\mathbf{I}$ | unit tensor |
| $J$ | heat source fluence ($J/m^2$) |
| $k$ | thermal conductivity [ W/m-K ] |
| $k_B$ | Boltzmann constant [J/K] |
| $l$ | thickness [m] |
| L | dimensionless thickness |
| $M$ | molar mass [kg/kmol] |
| $N$ | number density of atom [$m^{-3}$] |
| P | pressure [Pa] |
| $\mathbf{q}''$ | heat flux vector [$W/m^2$] |
| $R$ | reflectivity |
| $R_g$ | gas constant [J/kg-K] |
| s | solid-liquid interfacial location [m] |
| $S$ | intensity of the internal heat source [ $W/m^3$ ] |
| $t$ | time [ $s$ ] |
| $t_p$ | FWHM (Full Width at Half Maximum) pulse width [ s ] |
| $T$ | temperature [ K ] |
| $T_m$ | melting point [K] |
| $T_{l,I}$ | lattice temperature at solid-liquid interface [K] |
| $u_s$ | solid-liquid interfacial velocity [m/s] |
| $\mathbf{u}$ | electron velocity vector [m/s]<br>displacement vector of metal lattice [m] |
| $U$ | dimensionless displacement |
| $\mathbf{v}$ | mean (drift) velocity vector [m/s] |
| $V_0$ | interfacial velocity factor [m/s] |
| $x$ | coordinate [m] |

*Greek Symbols*

| | |
|---|---|
| $\alpha$ | thermal diffusivity [$m^2/s$] |
| $\beta$ | dimensionless time |
| $\delta$ | optical penetration depth [m] |
| $\delta_b$ | ballistic range [m] |
| $\eta$ | thermomechanical coupling factor |
| $\kappa_\varepsilon$ | thermal expansion coefficient of strain |
| $\lambda$ | Lamé constant |
| $\mu$ | Lamé constant |
| $\nu$ | Poisson's ratio |
| $\theta$ | dimensionless temperature |
| $\rho$ | density [ $kg/m^3$ ] |

$\tau$ relaxation time [s]

$\tau_q$ phase lag for the heat flux vector [s]

$\tau_T$ phase lag for the temperature gradient [s]

$\xi$ electron energy density [J]

dimensionless coordinate

*Subscripts*

$e$ electron

F Fermi

$i$ initial

$l$ lattice

$\ell$ liquid

$s$ solid